\documentclass[
preprint,
amsmath,amssymb, aps, prd, showpacs, nofootinbib ]{revtex4-1}
\usepackage{amsmath, amssymb, graphics}
\usepackage{graphicx}
\usepackage{subfigure}
\usepackage{float}
\usepackage{textcomp}
\usepackage{slashed}
\usepackage{hyperref}
\usepackage[mathlines]{lineno}
\usepackage{color}
\usepackage{mathrsfs}
\usepackage{makecell}
\newcommand{\ds}{\displaystyle×}
\newcommand{\comment}[1]{}
\usepackage{def}

\newcommand{\umassAff}{\affiliation{
    Amherst Center for Fundamental Interactions,
    Department of Physics, \\
    University of Massachusetts,
    Amherst,
    MA 01003, USA
}}

\newcommand{\TDLIAff}{\affiliation{
Tsung-Dao Lee Institute and  School of Physics and Astronomy, Shanghai Jiao Tong University, 800 Dongchuan Road, Shanghai, 200240 China
}}

\newcommand{\CaltechAff}{\affiliation{Kellogg Radiation Laboratory, California Institute of Technology, Pasadena, CA 91125 USA
}}

\newcommand{\itpAff}{\affiliation{
CAS Key Laboratory of Theoretical Physics, Institute of Theoretical Physics, Chinese Academy of Sciences, Beijing 100190, P. R. China
}}

\begin{document}

\hfill{ACFI T19-03}

\title{{ Probing a Scalar Singlet-Catalyzed Electroweak Phase Transition with Resonant Di-Higgs Production in the $4b$ Channel}}
\author{Hao-Lin Li}
\umassAff
\itpAff
\email{haolinli@itp.ac.cn}
\author{Michael J. Ramsey-Musolf}
\email{mjrm@physics.umass.edu}
\umassAff
\TDLIAff
\CaltechAff

\author{St\'ephane Willocq}
\email{willocq@physics.umass.edu}
\umassAff

\begin{abstract}
{We investigate the prospective reach of the 14 TeV HL-LHC for resonant production of a heavy Higgs boson that decays to two SM-like Higgs bosons in the $4b$ final state in the scalar singlet extended Standard Model. We focus on the reach for choices of parameters yielding a strong first order electroweak phase transition. 
The event selection follows the $4b$ analysis by the ATLAS Collaboration, enhanced with the use of
a boosted decision tree method to optimize the discrimination between signal and background events.
The output of the multivariate discriminant is used directly in the statistical analysis.  
The prospective reach of the $4b$ channel is compatible with previous projections for the $bb\gamma\gamma$ and $4\tau$ channels for heavy Higgs boson mass $m_2$ below 500 GeV and superior to these channels for $m_2 > 500$ GeV. With 3 ab$^{-1}$ of integrated luminosity, it is possible to discover the heavy Higgs boson in the $4b$ channel for $m_2< 500$ GeV in regions of parameter space yielding a strong first order electroweak phase transition and satisfying all other phenomenological constraints.}
\end{abstract}
\maketitle
\section{Introduction\label{sec:introduction}}
{After the discovery of the Higgs boson at the Large Hadron Collider (LHC)~\cite{Aad:2012tfa,Chatrchyan:2012xdj}, understanding the details of electroweak symmetry-breaking (EWSB) in the context of the thermal history of the universe
remains an important challenge for particle physics. In particular, it is possible that EWSB was accompanied by generation of the cosmic baryon asymmetry if new physics beyond the Standard Model (BSM) was active during that era.
The {\sc Planck} measurement of this asymmetry, characterized by the baryon-to-entropy density ratio $Y_B=n_b/s$,
gives~\cite{Ade:2013zuv}:
\begin{equation}
Y_B=(8.59\pm 0.11)\times10^{-11}\ .
\end{equation}
Explaining the origin and magnitude of $Y_B$ is a key problem for BSM scenarios. }Electroweak baryogenesis (EWBG) is one of the appealing { possibilities, in part due to its linking of $Y_B$ to EWSB and in part} due to its testability in current and near future experiments. Three ``Sakharov conditions''~\cite{Sakharov:1967dj} need  to be satisfied for a successful EWBG: baryon number (B) violation, C and CP violation, and departure from thermal equilibrium (through a strong first order electroweak phase transition) or a breakdown of CPT symmetry. 
In the Standard Model (SM), the first condition -- baryon number -- violation can be induced by the process of electroweak sphalerons. However, the CP violation in the SM is too feeble, and the EWSB transition is a crossover transition given the observed SM Higgs mass $m_h\sim 125$ GeV~\cite{Aoki:1999fi,Csikor:1998eu,Laine:1998jb,Gurtler:1997hr,Kajantie:1996mn}. Therefore, the minimal SM cannot generate a successful strong first order electroweak phase transition (SFOWEPT). {On the other hand, if new scalars exist in addition to the SM Higgs doublet, their interactions with the SM Higgs doublet may catalyze a SFOEWPT, thereby providing the necessary conditions for successful EWBG.\footnote{ New CP-violating interactions would also be required, a topic we do not treat further here.}}

In this paper, we focus on the singlet extension to the SM, the xSM, which is proven to be able to give a SFOEWPT~\cite{Profumo:2007wc,Espinosa:2011ax}. In the xSM, after EWSB, the gauge eigenstates of the singlet scalar and the SM Higgs doublet mix with each other to form the mass eigenstates $h_1$ (SM-like) and $h_2$ (singlet-like). 
Further, we restrict our study to 
searching for a signal of the on-shell production of the heavy singlet-like Higgs $h_2$ decaying into two SM-like Higgs $h_1$ (i.e. $m_2>2m_1$), because the regions of parameter space that can generate SFOEWPT simultaneously tend to enhance the $h_2h_1h_1$ tri-linear couplings~\cite{Profumo:2007wc,Espinosa:2011ax,No:2013wsa}.
Currently, the ATLAS and CMS experiments are searching for a resonant di-Higgs signal through different Higgs decay final states: $4b$~\cite{Aaboud:2018knk,Sirunyan:2018zkk}, $bb WW^*$ or $bbZZ^*$~\cite{Aaboud:2018zhh,Sirunyan:2017guj}, $bb\tau\tau$~\cite{Aaboud:2018sfw,Sirunyan:2017djm}, $bb\gamma\gamma$~\cite{Aaboud:2018ftw,Sirunyan:2018iwt}, $WW^* WW^*$~\cite{Aaboud:2018ksn}, and $\gamma\gamma WW^*$~\cite{Aaboud:2018ewm}. Thus far, no significant excess over SM backgrounds has been observed.
On the theoretical side, several studies have been performed in the parameter regions that are viable for SFOEWPT. The singlet-like $h_2$ with a relatively light mass ($\sim$270 GeV) can be discovered in the $bb\tau\tau$ final state at the 14 TeV LHC with a luminosity of 100 fb$^{-1}$~\cite{No:2013wsa}. In the $bb\gamma\gamma$ and $4\tau$ final states, a discovery is possible for $m_2$ up to 500 GeV at the 14 TeV high-luminosity LHC (HL-LHC) with a luminosity of 3~ab$^{-1}$~\cite{Kotwal:2016tex}. In the $bbWW^*$ final state, a resonant signal can be discovered for $m_2$ in the range between 350 GeV and 600 GeV at the 13 TeV LHC with a luminosity of 3~ab$^{-1}$~\cite{Huang:2017jws}. 

In this paper, we study the prospective discovery/exclusion in the $4b$ final state at the 14 TeV HL-LHC with a luminosity of 3~ab$^{-1}$. To that end, we first identify 22 benchmark points with $m_2\in[300, 850]$~GeV that produce the maximal and minimal di-Higgs signal rate $\sigma_{h_2}\times {\rm BR}(h_2\to h_1h_1)$ in consecutive 50 GeV intervals. The selected benchmark points satisfy all the current phenomenological constraints from the Higgs signal rate and electroweak precision data, and also satisfy the theoretical constraints from vacuum stability, perturbativity, and a SFOEWPT. We perform a full simulation of signal and background processes with the MadGraph5 parton level event generator~\cite{Alwall:2014hca} using PYTHIA6~\cite{Sjostrand:2006za} to simulate the parton shower and the DELPHES3 fast detector simulation~\cite{deFavereau:2013fsa}. Further, we use the TMVA package~\cite{TMVA} to implement the Boosted Decision Tree (BDT) algorithm to optimize the event selection, finally obtaining the signal significance from the BDT score distributions of background and signal events.

Based on this analysis and the results shown in Fig.~\ref{fig:14BMsig} below, we arrive at the following conclusions:
\begin{itemize}
\item For singlet-like Higgs masses below 500 GeV, the significance of the $4b$ final state is competitive with the $bb\gamma\gamma$ and $4\tau$ final states, and it is possible to make a discovery at the 14 TeV HL-LHC with a luminosity of 3 ab$^{-1}$ for some portions of the SFOEWPT-viable parameter space.
\item For singlet-like Higgs masses above 500 GeV, the significance of the $4b$ final state is higher than in the $bb\gamma\gamma$ and $4\tau$ final states but somewhat below recent projections for the $bbWW^\ast$ final state.
\item With the results of the benchmark models that produce minimal di-Higgs signal rate, we found that it is impossible to exclude (at the 95\% confidence level)  all portions of parameter space consistent  with a SFOEWPT and present phenomenological constraints at the HL-LHC.
\end{itemize}

The discussion of our analysis leading to these conclusions is organized as follows: Sec.~\ref{sec:model} introduces the xSM framework and describes both theoretical and phenomenological constraints. In Sec.~\ref{sec:EWPT}, we describe the requirements for a SFOEWPT and the parameter scan. In Sec.~\ref{sec:analysis}, we discuss the simulation and analysis of the $4b$ signal and background in detail and also present prospects for the 14 TeV HL-LHC. Section~\ref{sec:concl} is dedicated to the conclusions. In the Appendix, we perform a global analysis of ATLAS Run~2 single Higgs measurements and present the distributions of the kinematic variables used in the BDT analysis.

\section{The xSM\label{sec:model}} 

\subsection{The Model}
The most general, renormalizable scalar potential in the xSM model is given by:
\begin{eqnarray}
\label{ScalarPotential1}
 V(H,S) =\ds  -\mu^2 \left( H^\dagger H \right) + \lambda \left( H^\dagger H \right)^2 + \frac{a_1}{2} \left( H^\dagger H \right) S & \nonumber \\
 \ds + \frac{a_2}{2} \left( H^\dagger H \right) S^2 + \frac{b_2}{2} S^2 + \frac{b_3}{3} S^3 + \frac{b_4}{4} S^4 , &
\label{TotPot}
\end{eqnarray}
where $S$ is the real singlet and $H$ is the SM Higgs doublet. When $S$ obtains a vacuum expectation value (vev, see below), the $a_1$ and $a_2$ parameters induce  mixing between the singlet scalar and the SM Higgs doublet, thereby providing a portal for the singlet scalar to interact with other SM particles. A $\mathbb{Z}_2$ symmetry is present in the absence of $a_1$ and $b_3$ { terms, a necessary condition for $S$ to be a viable dark matter candidate. In what follows, however, we retain both parameters in our study as they play an important role in the strength of the electroweak phase transition (EWPT) and also in the di-Higgs signal rate at collider experiments.

After EWSB, $H\to (v_0 +h)/\sqrt{2}$ with $v_0=246$ GeV, and $S\to x_0+s$ where $x_0$ is the vev for $S$ without loss of generality. The stability of the scalar potential requires the quartic coefficients along all the directions in the field space to be positive. {This translates into a requirement of a positive Hessian determinant of the potential with respect to fields $s$ and $h$:
\begin{eqnarray}
{\rm det}\left(\begin{array}{c c}
\partial^2 V/(\partial s^2) & \partial^2 V/(\partial s \partial h) \\
\partial^2 V/(\partial h \partial s) & \partial^2 V/(\partial h^2)
\end{array}\right)>0.
\end{eqnarray}
}
This leads the bounds $\lambda>0$, $b_4 >0$ and $a_2>-2\sqrt{\lambda b_4}$.
Another way to obtain these bounds is by parameterizing ($h$,$s$) as ($r\cos\alpha$, $r\sin\alpha$) in the field space, and we are able to extract the quartic coefficients of $r$ along the $\alpha$ direction in the field space:
\begin{eqnarray}
\frac{1}{4}\left( (b_4+\lambda - a_2)\cos ^4 \alpha + (a_2-2b_4)\cos^2 \alpha +b_4\right).
\end{eqnarray}
Requiring the above expression be larger than zero for any value of $\cos\alpha$ also leads to the same conditions. 

Utilizing the minimization conditions, 
\begin{eqnarray}
\label{minimization of potential}
\left. \frac{dV}{dh}\right\vert_{h=0,s=0}=0,\quad \left. \frac{dV}{ds}\right\vert_{h=0,s=0}=0,
\end{eqnarray}
one can express two potential parameters in Eq.~(\ref{ScalarPotential1}) in terms of the vevs and other parameters:
\begin{eqnarray}
& \ds \mu^2 = \lambda v_0^2 + \left( a_1 + a_2 x_0 \right) \frac{x_0}{2}, & \nonumber \\
& \ds b_2 = \ds - b_3 x_0 - b_4 x_0^2 - \frac{a_1 v_0^2}{4 x_0} - \frac{a_2 v_0^2}{2} .&
\label{eq:ewsb}
\end{eqnarray}
Two additional conditions need to be satisfied for ($v_0$, $x_0$) to be a stable minimum. One of them is that ($v_0$, $x_0$) minimizes the potential locally, implying that:

\begin{eqnarray}
\ds b_3 x_0 + 2 b_4 x_0^2 - \frac{a_1 v_0^2}{4 x_0} - \frac{ (a_1 + 2 a_2 x_0 )^2 }{ 8 \lambda } > 0 .
\end{eqnarray}
Also, this minimum point should be a global minimum, a requirement that we impose numerically.

{As for the perturbativity consideration, we have the following na\"ive requirements on the quartic couplings}: 
\begin{eqnarray}
\left| \frac{a_1}{2}\right|,\left|\frac{a_2}{2}\right|,\left|\frac{b_4}{4}\right| < 4\pi .
\end{eqnarray}
However, as discussed in Sec.~\ref{sec:EWPT} when scanning over the parameter space for benchmark points we implement more stringent bounds on those parameters compared with the above requirements. One may refer to Refs.~\cite{Robens:2015gla,Robens:2016xkb,Gonderinger:2009jp} for more details about the perturbativity bound in the xSM.

Now we obtain the elements of the mass-squared matrix by:
\begin{eqnarray}
& m_{h}^2 \equiv \ds \frac{d^2 V}{dh^2} = 2 \lambda v_0^2 , & \nonumber \\
& m_{s}^2 \equiv \ds \frac{d^2 V}{ds^2}  = b_3 x_0 + 2 b_4 x_0^2 - \frac{a_1 v_0^2}{4 x_0} , & \nonumber \\
& m_{hs}^2 \equiv \ds \frac{d^2 V}{dh ds} = \left(a_1 + 2 a_2 x_0 \right) \frac{v_0}{2} .
\label{mixingM}
\end{eqnarray}
After the diagonalization of the above mass matrix, the physical masses of two neutral scalars can be expressed as:
\begin{eqnarray}
& m_{2,1}^2 = \ds \frac{ m_{h}^2 + m_{s}^2 \pm \left| m_{h}^2 - m_{s}^2 \right| \sqrt{ 1 + \ds \left( \frac{4 m_{hs}^2 }{ m_{h}^2 - m_{s}^2 } \right)^2 } } {2} ,& \nonumber \\
\label{Meigenvalues}
\end{eqnarray}
with $m_2>m_1$ by construction. The mass eigenstates and gauge eigenstates are related by a rotation matrix:

\begin{eqnarray}
\begin{pmatrix} 
  h_1    \\ 
  h_2 
\end{pmatrix}
=
\begin{pmatrix} 
  \cos\theta & \sin\theta    \\ 
  -\sin\theta & \cos\theta 
\end{pmatrix}
\begin{pmatrix} 
  h    \\ 
  s 
\end{pmatrix},
\label{mixingmatrix}
\end{eqnarray}
where $h_1$ is the SM-like Higgs boson with $m_1=125$~GeV, and $h_2$ is identified as the  singlet-like mass eigenstate. The mixing angle $\theta$ can be expressed in terms of the vevs, physical masses and potential parameters:
\begin{eqnarray}
\sin 2 \theta =  \ds \frac{ 2 m_{hs}^2 }{ \left( m_1^2 - m_2^2 \right) } = \ds \frac{ \left( a_1 + 2 a_2 x_0 \right) v_0  }{ \left( m_1^2 - m_2^2 \right) } .
\label{sin2theta}
\end{eqnarray}

From Eq.~(\ref{mixingmatrix}), one can observe that the couplings of $h_1$ and $h_2$ to the SM vector bosons and fermions are rescaled with respect to their SM Higgs couplings:
\begin{eqnarray}
g_{h_1 XX} = \cos \theta \; g_{h XX}^{\mathrm{SM}}, \quad g_{h_2 XX} = \sin \theta \; g_{h XX}^{\mathrm{SM}},
\end{eqnarray}
where $XX$ represents final states consisting of pairs of SM vector bosons or fermions.
In this case, all the signal rates associated with the single Higgs measurements are rescaled by the mixing angle only:
\begin{eqnarray}
\mu_{h_1 \to XX} = \frac{\sigma_{h_1} \cdot \text{BR}}{\sigma^\mathrm{SM}_{h_1} \cdot \text{BR}^{\mathrm{SM}}} = \cos^2 \theta ,
\end{eqnarray} 
where $\sigma_{h_1}$ and BR are the production cross section and branching ratio in the xSM, and the quantities with the superscript SM are the corresponding values in the SM. In the xSM for $m_2>m_1$, we have BR = $\text{BR}^\mathrm{SM}$ because the partial width of each decay mode is rescaled by $\cos ^2\theta$ and there is no new decay channel appearing.

In order to investigate the di-Higgs production, we also require the tri-Higgs couplings. The one relevant for the resonant di-Higgs production is $\lambda_{211}$: 
\begin{eqnarray}
\label{g211}
\lambda_{211} = \frac{1}{4}\left[ (a_1+2a_2x_0) \cos^3 \theta + 4 v_0 (a_2 -3 \lambda) \cos^2 \theta \sin \theta \right. \nonumber\\
\left. + (a_1+2a_2x_0 -2b_3-6b_4x_0) \cos \theta \sin^2 \theta -2a_2 v_0 \sin^3 \theta \right]. 
\end{eqnarray}

In this work, we focus on the situation where $m_2>2m_1$ such that a resonant production of $h_2$ and a subsequent decay to $h_1h_1$ is allowed. Therefore, we are able to calculate the partial width $\Gamma_{h_2\to h_1 h_1}$:
\begin{eqnarray}
\Gamma_{h_2 \to h_1 h_1} = \frac{\lambda_{211}^2 \sqrt{1 - \ds 4 m_1^2 / m_2^2 } }{8 \pi m_2} , 
\label{partialWidthh1h1}
\end{eqnarray}
and the total width of $h_2$:
\begin{eqnarray}
\Gamma_{h_2} = \mathrm{sin}^2 \theta \; \Gamma^{\mathrm{SM}}(m_2) + \Gamma_{h_2 \to h_1 h_1} ,
\end{eqnarray} where $\Gamma^{\mathrm{SM}}(m_2)$ represents the total width of the SM Higgs boson with a mass of $m_2$, which is taken from Ref.~\cite{Heinemeyer:2013tqa}.
The signal rate for $pp\to h_2\to XX$ 
normalized to the SM value is given by:
\begin{eqnarray}
\mu_{h_2 \to XX} = \sin^2 \theta \left( \frac { \sin^2 \theta \; \Gamma^{\mathrm{SM}}(m_2)}{ \Gamma_{h_2} }  \right),
\end{eqnarray}
which will be used to constrain the parameter space in the next section. The production cross section for the process $pp\to h_2\to h_1h_1$ can also be calculated:
\begin{eqnarray}
\sigma_{h_1h_1} = \sigma^\mathrm{SM}(m_2)\times s^2_\theta\frac{\Gamma_{h_2\to h_1h_1}}{s^2_{\theta}\Gamma^\mathrm{SM}(m_2)+\Gamma_{h_2\to h_1 h_1}} \ ,
\end{eqnarray}
{where $s_\theta\equiv \sin\theta$ and, for future reference, $c_\theta\equiv \cos\theta$}.

\subsection{Phenomenological Constraints on the Model Parameters}

The mixing angle $\theta$ between the singlet and the SM Higgs doublet in the xSM is constrained by measurements of the single SM-like Higgs signal strengths.
We obtain a 95\% C.L. upper limit on $\sin^2\theta$ of 0.131 by performing a global fit with current ATLAS Run~2 single Higgs measurements as discussed in Appendix~\ref{sec:gf}.

The LHC searches for the heavy neutral Higgs boson also provide constraints on the parameter space. Here, we take into account the existing limits on both the $h_2\to VV$~\cite{Sirunyan:2018qlb,Aaboud:2018bun,Aad:2015kna,Aad:2015ipg,Chatrchyan:2013yoa,Khachatryan:2015cwa} and the $h_2\to h_1h_1$ decays, where $h_1h_1$ decay into 4$b$~\cite{Aaboud:2018knk,Sirunyan:2018zkk}, $b\bar{b}\gamma\gamma$~\cite{Aaboud:2018ftw,Sirunyan:2018iwt}, or $b\bar{b}\tau\tau$~\cite{Aaboud:2018sfw,Sirunyan:2017djm}. The constraints on the ($m_2$, $c_\theta$) plane can be found in our previous work~\cite{Huang:2017jws}.
We will also guarantee each benchmark point in the parameter scan in the next section satisfies all the limits mentioned above.

Finally, we discuss the constraints from electroweak precision observables (EWPO). The mixing between the singlet scalar and the SM Higgs doublet induces modifications of the oblique parameters $S$, $T$, and $U$ with respect to their SM values. From Eq.~(\ref{mixingmatrix}), the deviation in oblique parameters ${\cal O}$, denoted by $\Delta {\cal O}$, can be expressed in terms of the SM Higgs contribution to that parameter, ${\cal O}^\mathrm{SM}(m)$~\cite{Peskin:1991sw,Hagiwara:1994pw} and the mixing angle $\theta$, where $m$ is either $m_1$ or $m_2$:
\begin{eqnarray}
\label{oblique_1}
\Delta \mathcal{O} = (c^2_{\theta} -1) \mathcal{O}^{\mathrm{SM}}(m_1) + s^2_{\theta} \; \mathcal{O}^{\mathrm{SM}}(m_2) 
= s^2_{\theta} \left[ \mathcal{O}^{\mathrm{SM}}(m_2) - \mathcal{O}^{\mathrm{SM}}(m_1) \right] \, . 
\end{eqnarray}
In the xSM, the parameter $U=0$ is a good approximation; we therefore focus only on the deviations in the $S$ and $T$ parameters, which we take from the Gfitter group~\cite{Baak:2014ora}:
\begin{equation}
 \label{chi_EWPO1}
\begin{array}{c}
\Delta S \equiv S - S_{\mathrm{SM}} = 0.06 \pm 0.09 \\
\Delta T \equiv T - T_{\mathrm{SM}} = 0.10 \pm 0.07
\end{array} \quad \quad \quad \quad 
\rho_{ij} = \left(\begin{array}{cc}
             1 & 0.91\\
             0.91 &1
            \end{array}\right),
\end{equation}
where $\rho_{ij}$ is the covariance matrix in the ($S$,$T$) plane. Again, we will impose the criteria in the parameter scan in the next section such that for each benchmark point, $\Delta \chi^2 (m_2, c_\theta)$ defined below is less than 5.99, which corresponds to deviations of $S$ and $T$ parameters within 95\% C.L.:
\begin{equation}
 \label{chi_EWPO2}
\Delta\chi^2(m_2,c_{\theta}) = \sum_{i,j} \left[\Delta\mathcal{O}_{i}(m_2,c_{\theta}) - \Delta\mathcal{O}^0_{i}\right] (\sigma^2)_{ij}^{-1}
\left(\Delta\mathcal{O}_{j}(m_2,c_{\theta}) - \Delta\mathcal{O}^0_{j}\right)\, ,
\end{equation}
where the $\Delta\mathcal{O}^0_{i}$ denote the central values in Eq.~(\ref{chi_EWPO1}) and $(\sigma^2)_{ij} \equiv \sigma_i \rho_{ij} \sigma_j$, with $\sigma_i$ being the error in $S$ or 
$T$ as indicated in Eq.~(\ref{chi_EWPO1}). 
One can observe from Fig.~1 in Ref~\cite{Huang:2017jws} that in general the upper limit for $\sin^2\theta$ extracted from EWPO is more stringent than the bound obtained from the Higgs global fit, with a limit changing from 0.12 for $m_2 = 250$~GeV to 0.04 for $m_2 = 950$~GeV.

\section{Electroweak Phase Transition and Benchmarks for Di-Higgs Production}\label{sec:EWPT}
The character of EWPT is understood in terms of the finite-temperature effective potential, $V^{T\neq 0}_{eff}$ However, the fact that the standard derivation of $V^{T\neq 0}_{eff}$ suffers from gauge dependence is well known which is discussed in depth in Ref.~\cite{Quiros:1994dr}. Here we employ a high-temperature expansion to restore the gauge independence in our analysis (see Ref.~\cite{Profumo:2014opa} for details). {In such a case, we include in our finite temperature effective potential the $T=0$ tree level potential and the gauge-independent thermal mass corrections to $V^{T\neq 0}_{eff}$, which are crucial to restore electroweak symmetry at high temperature.} In this limit, the $a_1$ and $b_3$ parameters  {will generate a tree-level barrier between the broken and unbroken electroweak phases, thereby allowing for a first-order EWPT. We also note that the presence of the $a_2$ term may also strengthen the first order transition, as discussed in Ref.~\cite{Profumo:2007wc}.}
 In the high-temperature limit, we follow Ref.~\cite{Pietroni:1992in,Profumo:2007wc} 
 and write the $T$-dependent, gauge-independent (indicated by the presence of a bar) vevs in a cylindrical coordinate representation as
\begin{equation}
\bar{v}(T) / \sqrt{2} = \bar{\phi}(T) \cos \alpha (T), \quad \bar{x}(T) = \bar{\phi}(T) \sin \alpha (T),
\end{equation}
with $\bar{v}(T=0)=v_0$ and $\bar{x}(T=0)=x_0$. The critical temperature $T_c$ is defined as the temperature at which the broken and unbroken phases are degenerate, i.e. $V^{T\neq 0}_{eff}(\phi,\alpha\neq\pi/2, T_c)=V^{T\neq 0}_{eff}(\phi,\alpha=\pi/2, T_c)$. Once the critical temperature is found, one is able to evaluate the quenching effect of the sphaleron transitions in the broken electroweak phase (see, e.g., Ref.~\cite{Morrissey:2012db}), which is related to the energy of the electroweak sphaleron that is proportional to the vev of SU(2)$_L$ doublet $\bar{v}(T)$. A first-order EWPT is strong when the quenching effect is sufficiently large, and the criterion is approximately given by:
\begin{equation}
\cos \alpha (T_c ) \,\frac{\bar{\phi} (T_c)}{T_c} \gtrsim 1.
\end{equation}

To select the benchmarks parameter points for the collider simulation, we perform a scan over the parameters $a_1$, $b_3$, $x_0$, $b_4$, and $\lambda$ within the following ranges:
\begin{equation}
a_1/\text{TeV}, \quad b_3/\text{TeV} \in [-1, 1], \quad x_0/\text{TeV} \in [0, 1], \quad b_4, \lambda \in [0, 1],
\end{equation}  
while the remaining parameters are fixed from the input values of $v_0=246$ GeV and $m_h=125$ GeV. Our lower bound on quartic couplings $b_4$ and $\lambda$ guarantees tree-level vacuum stability. We also require a na\"ive perturbativity bound on the Higgs portal coupling $a_2/2\lesssim 5$. For each set of randomly chosen parameters, we calculate $c_\theta$, $m_2$, and $\lambda_{211}$, and only keep the points that satisfy all the phenomenological constraints mentioned in the previous section (Higgs signal rate, LHC search for heavy Higgs $h_2$, and EWPO). We then pass these sets of parameters into the {\sc CosmoTransitions} package~\cite{Wainwright:2011kj} and numerically evaluate all the quantities related to the EWPT, such as critical temperature, sphaleron energy, tunneling rate into the electroweak symmetry-broken phase, using as an input the xSM finite temperature effective potential  in the high-temperature limit. Finally, we only keep the sets of parameters that satisfy the strong first-order EWPT criterion defined above and also have a sufficient tunneling rate to prevent the universe from remaining in a metastable vacuum.

From the randomly chosen parameters satisfying the foregoing requirements, we identify benchmark points with maximum and minimum signal rate $\sigma(pp\to h_2) \times {\rm BR}(h_2 \to h_1 h_1)$ from 11 consecutive $h_2$ mass windows of width 50 GeV ranging from 300 to 850 GeV. The upper bound of $m_2=850$ GeV is obtained by the observation that we did not find a choice of parameters for $m_2$ larger than 850 GeV that give a SFOEWPT, even though our scan in $m_2$ reaches one TeV. We list all the benchmark points in Tables~\ref{tab:T1} and ~\ref{tab:T2}. We would like to mention that the benchmark points B3 and B4 for maximum signal rate in Table~\ref{tab:T1} has already been excluded by the CMS $h_2\to ZZ$ search~\cite{Sirunyan:2018qlb}, but we retain them here to make contact with the results of previous studies for comparison. {In contrast, the new ATLAS and CMS limits on resonant di-Higgs production in the $b{\bar b} \tau \tau$ channel\cite{Aaboud:2018sfw,Sirunyan:2017djm} do not yet appear to constrain the SFOEWPT-viable parameter space}.

\begin{table*}[t!]
\resizebox{\columnwidth}{!}{%
  \begin{tabular}{ c  c  c  c  c  c  c  c  c c c c  c c c }
    \hline
    \hline
 Benchmark & $\cos \theta$ & $m_2$ & $\Gamma_{h_2}$ & $x_0$ & $\lambda$ & $a_1$ & $a_2$ & $b_3$ & $b_4$ & $\lambda_{111}$ & $\lambda_{211}$ & $\sigma$ & BR ~ \\         
           &               & (GeV) &     (GeV)        & (GeV) &           & (GeV) &       & (GeV) &       &      (GeV)       &       (GeV)     &  ~  (pb) ~  &  \\         
    \hline
B1 & 0.974 &  327 & 0.929 &  60.9 & 0.17 & -490 & 2.65 & -361 & 0.52 & 45 & 62.2 & 0.56 & 0.33 \\
B2 & 0.980 &  362& 1.15 &  59.6 & 0.17 & -568 & 3.26 & -397 & 0.78 & 44.4 & 76.4 & 0.48 & 0.40 \\
B3 & 0.983 &  415 & 1.59 &  54.6 & 0.17 & -642 & 3.80 & -214 & 0.16 & 44.9 & 82.5 & 0.36 & 0.33 \\
B4 & 0.984 &  455 & 2.08 &  47.4 & 0.18 & -707 & 4.63 & -607 & 0.85 & 46.7 & 93.5 & 0.26 & 0.31 \\
B5 & 0.986 &  511 & 2.44 &  40.7 & 0.18 & -744 & 5.17 & -618 & 0.82 & 46.6 & 91.9 & 0.15 & 0.24 \\
B6 & 0.988 &  563 & 2.92 &  40.5 & 0.19 & -844 & 5.85 & -151 & 0.08 & 47.1 & 104 & 0.087 & 0.23 \\
B7 & 0.992 &  604 & 2.82 &  36.4 & 0.18 & -898 & 7.36 & -424 & 0.28 & 45.6 & 119 & 0.045 & 0.30 \\
B8 & 0.994 &  662 & 2.97 &  32.9 & 0.17 & -976 & 8.98 & -542 & 0.53 & 44.9 & 132 & 0.023 & 0.33 \\
B9 & 0.993 &  714 & 3.27 &  29.2 & 0.18 & -941 & 8.28 & 497 & 0.38 & 44.7 & 112 & 0.017 & 0.20 \\
B10 & 0.996 &  767 & 2.83 &  24.5 & 0.17 & -920 & 9.87 & 575 & 0.41 & 42.2 & 114 & 0.0082 & 0.22 \\
B11 & 0.994 &  840 & 4.03 & 21.7 & 0.19 & -988 & 9.22 & 356 & 0.83 & 43.9 & 83.8 & 0.0068 & 0.079 \\
 \hline
 \hline
  \end{tabular}
  }
  \caption{\small Values of the various xSM independent and dependent parameters for each of the benchmark values consistent with a SFOEWPT chosen to {\bf maximize} 
  the $\sigma(pp \to h_2)\times\mathrm{BR}(h_2\to h_1h_1)$ value at the 14~TeV LHC.}
\label{tab:T1}
\end{table*} 

\begin{table*}[t!]
  \begin{tabular}{ c  c  c  c  c  c  c  c c c c  c c c  c }
    \hline
    \hline
 Benchmark & ~ $\cos \theta$ ~  & ~ $m_2$ ~ & ~ $\Gamma_{h_2}$ ~  & ~ $x_0$ ~ & ~ $\lambda$ ~ & ~ $a_1$ ~ & ~ $a_2$ ~ & ~ $b_3$ ~ & ~ $b_4$ ~ & ~ $\lambda_{111}$ ~ & ~ $\lambda_{211}$ ~ & ~ $\sigma$ ~  & ~ BR ~ \\         
           &               &               &     (GeV)        & (GeV) &           & (GeV) &       & (GeV) &       &      (GeV)       &       (GeV)     &  ~  (fb) ~  &  \\         
    \hline
BM1 & 0.9999 & 329 & 0.00593 &  111 & 0.13 & -812 & 3.61 & -99.8 & 0.35 & 31.8 & 7.30 & 1.1 & 0.71 \\
BM2 & 0.9995 & 363 & 0.0549 &  80.6 & 0.13 & -699 & 4.16 & -91.5 & 0.57 & 32.2 & 21.6 & 8.2 & 0.68 \\
BM3 & 0.9803 & 419 & 1.32 &  234 & 0.18 & -981 & 1.56 & 0.417 & 0.96 & 39.0 & 17.5 & 6.9 & 0.018 \\
BM4 & 0.9997 & 463 & 0.0864  & 56.9 & 0.13 & -763 & 6.35 & 113 & 0.73 & 32.2 & 27.4 & 3.0 & 0.63 \\
BM5 & 0.9994 & 545 & 0.278 &  50.3 & 0.13 & -949 & 8.64 & 152 & 0.57 & 33.0 & 51.6 & 2.9 & 0.62 \\
BM6 & 0.9991 & 563 & 0.459 &  33.0 & 0.13 & -716 & 9.25 & -448 & 0.96 & 33.7 & 66.8 & 3.7 & 0.62 \\
BM7 & 0.9836 & 609 & 4.03 &  34.2 & 0.22 & -822 & 4.53 & -183 & 0.57 & 47.8 & 45.2 & 2.2 & 0.030 \\
BM8 & 0.9870 & 676 & 4.48 &  30.3 & 0.22 & -931 & 5.96 & -680 & 0.43 & 48.4 & 55.2 & 1.3 & 0.037 \\
BM9 & 0.9904 & 729 & 4.22 &	27.3 & 0.21 & -909 & 6.15 & 603 & 0.94 & 45.7 & 61.0 & 0.78 & 0.045 \\
BM10 & 0.9945 & 792 & 3.36 & 22.2 & 0.18 & -936 & 9.47 & -848 & 0.66 & 43.5 & 92.4 & 0.77 & 0.12 \\
BM11 & 	0.9944 & 841 & 3.95 & 21.2 & 0.19 & -955 & 8.69 & 684 & 0.53 & 43.3 & 73.4 & 0.28 & 0.062 \\
 \hline
 \hline
  \end{tabular}
  \caption{\small Values of the various xSM independent and dependent parameters for each of the benchmark values consistent with a SFOEWPT chosen to {\bf minimize} the $\sigma(pp \to h_2) \times \mathrm{BR}(h_2 \to h_1 h_1)$ at the 14~TeV LHC.}
\label{tab:T2}
\end{table*}    

\section{4b Final State Analysis}\label{sec:analysis}
\subsection{Reproduction of 13 TeV LHC results}\label{sec:13TeVanalysis}
For the signal process, the $h_2$ mass is varied from 300 GeV to 1500 GeV in steps of 100 GeV. For the background processes, we generate $pp\to 4b$ and $pp\to t\bar{t}$ with top quarks decaying hadronically. We follow the ATLAS resolved analysis in Ref.~\cite{Aaboud:2016xco}, and reproduce the signal efficiency and background distributions in Figs.~\ref{fig:13comp4b} and \ref{fig:distbg13}, respectively.
Parton level signal and background events are generated with MG5\_AMC@NLOv2.4.3~\cite{Alwall:2014hca} {and the NNPDF2.3QED LO set of parton distribution functions~\cite{Ball:2013hta}}. {For the $4b$ QCD background, we generate events with the process \texttt{p p > b b b$\sim$  b$\sim$}, while all other parton level cuts are set to the Madgraph default values. For the $t\bar{t}$ background, we generate events with the process \texttt{p p > t t$\sim$,(t > b c s$\sim$),(t$\sim$ > b$\sim$ c$\sim$ s)} plus one additional jet with jet matching. The \texttt{xqcut} in the run card is set to 20 GeV, and other cuts are kept at the default settings.} The events are interfaced with PYTHIA6~\cite{Sjostrand:2006za} for parton showering, fragmentation and hadronization. DELPHES3~\cite{deFavereau:2013fsa} is used to simulate the detector response. The default CMS DELPHES card is used rather than the ATLAS DELPHES card as it better approximates the $b$-tagging and jet reconstruction performance. Jets are constructed using the anti-$k_t$ clustering algorithm with a radius parameter $R$ set to 0.4, and the efficiency for a $b$-quark-initiated jet to pass the $b$-tagging requirements is {parameterized as a function of the jet transverse momentum $\pt$ in a manner corresponding to an average 70\% efficiency working point described in Ref.~\cite{Chatrchyan:2012jua}. (This is the default setting in the DELPHES CMS card).}

The selection criteria for the ATLAS analysis are as follows:
\begin{itemize}
\item Events must have at least four $b$-tagged jets with $\pt>$ 40 GeV and $|\eta|< 2.5$. If the number of $b$-tagged jets is greater than four, the four jets with the highest $\pt$ are selected to reconstruct two dijet systems in each event.
\item Two dijet systems are formed using the selected $b$-tagged jets. The two jets in each dijet system are required to have $\Delta R< 1.5$ and the transverse momentum of the leading (subleading) dijet system must be greater than 200 (150) GeV.
\item The leading and subleading dijet systems must satisfy the following set of requirements depending on the reconstructed invariant mass (\mfourj) of the four selected $b$-tagged jets:
\[\pt^{\mathrm{lead}} > \begin{cases}
		400\,{\rm GeV}& \mathrm{if}\ \mfourj > 910\,{\rm GeV}, \\
		200\,{\rm GeV}& \mathrm{if}\ \mfourj < 600\,{\rm GeV}, \\ 
		0.65\,\mfourj - 190\,{\rm GeV}& \mathrm{otherwise},
		\end{cases}
\]
\[\pt^{\mathrm{subl}} > \begin{cases}
	260\,{\rm GeV}& \mathrm{if }\ \mfourj > 990\,{\rm GeV}, \\
	150\,{\rm GeV}& \mathrm{if }\ \mfourj < 520\,{\rm GeV}, \\ 
	0.23\,\mfourj + 30\,{\rm GeV}& \mathrm{otherwise},
	\end{cases}
\]
\[ |\Delta\eta_{\rm dijets}| < \begin{cases}
	1.0 & \mathrm{if }\ \mfourj < 820\,{\rm GeV}, \\
	1.6\times10^{-3}\, \mfourj - 0.28 & \mathrm{otherwise}.
	\end{cases}
\]
\item To reduce the $t\bar{t}$ background, we impose a ``$t\bar{t}$ veto" as follows.
A set of $W$-boson candidates is formed by combining one of the $b$-tagged jets in the dijet system with any extra jet in the event that satisfies $\pt > 30$ GeV and $|\eta|< 2.5$ as well as $\Delta R < 1.5$ relative to the dijet system. Top-quark candidates are then formed by combining the dijet system with each of the extra jets that are selected. An event is vetoed if the invariant masses of the $W$-boson ($m_W$) and top-quark ($m_t$) candidates satisfy the following condition for any possible choice of extra jet and $b$-tagged jet from either of the dijet systems in the event:
\begin{equation}
X_{tt}\,=\,\sqrt{\left (\frac{m_{W}\,-\,80.4\,{\rm GeV}}{0.1\,m_{W}}\right )^2 + \left (\frac{m_{t}\,-\,172.5\,{\rm GeV}}{0.1\,m_{t}}\right )^2}~<3.2.
\end{equation}
\item Finally, the signal region is defined by the following requirement on the invariant masses of the leading and subleading dijet systems forming the two Higgs boson candidates:
\begin{equation}
X_{h_1h_1}\,=\,\sqrt{\left (\frac{\mtwoj^\mathrm{lead}\,-\,120\,{\rm GeV}}{0.1\,\mtwoj^\mathrm{lead}}\right )^2 +\, \left (\frac{\mtwoj^\mathrm{subl}\,-\,113\,{\rm GeV}}{0.1\,\mtwoj^\mathrm{subl}}\right )^2}~<1.6.
\end{equation} 
The central values for $\mtwoj^\mathrm{lead}$ and $\mtwoj^\mathrm{subl}$ in the above equation are somewhat lower than in the ATLAS analysis~\cite{Aaboud:2016xco} to account for differences in the treatment of jets in DELPHES compared to the ATLAS simulation.
\end{itemize} 
The acceptance times efficiency values for signal events with $m_2$ ranging from 500 to 1000 GeV are compared with the ATLAS results in Fig.~\ref{fig:13comp4b}.
Overall, the signal region efficiencies obtained in this analysis agree well with those from Ref.~\cite{Aaboud:2016xco}.

\begin{figure}[htp]
  \includegraphics[width=0.6\linewidth]{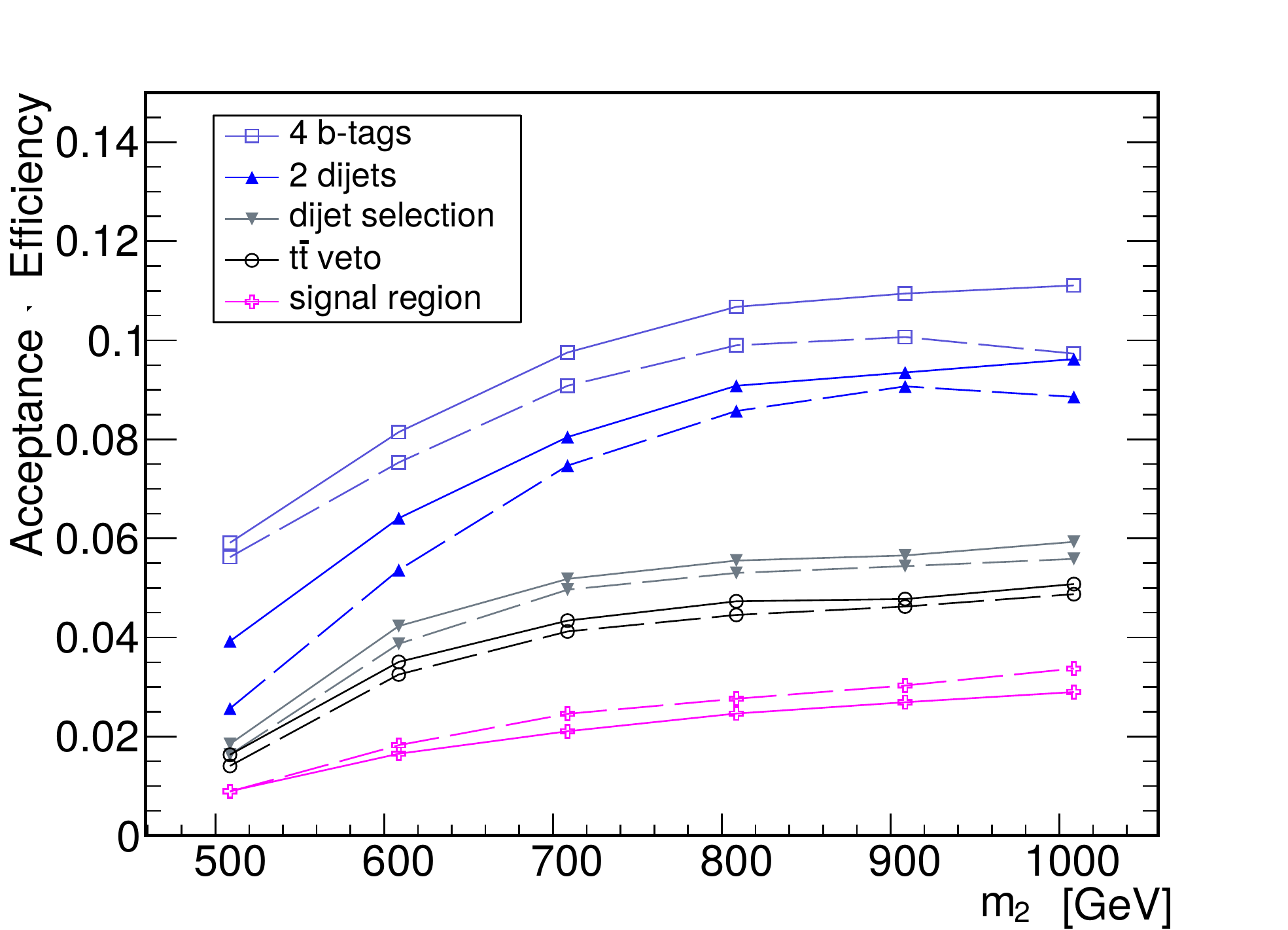}
  \caption{Acceptance times efficiency for signal events at successive stages of event selection for the 13 TeV HLC. Solid lines show the results of this analysis whereas dashed lines represent the ATLAS results from Ref.~\cite{Aaboud:2016xco}.
  }
  \label{fig:13comp4b}
\end{figure}

The background event yields in the signal region are summarized in Table~\ref{tab:13comp}. In addition to the yields from $4b$ and $t\bar{t}$ production, the contribution from $bbcc$ production with the $c$-quark jets passing the $b$-tagging requirements is estimated assuming that the kinematic distributions of jets in $bbcc$ events are similar to those of $4b$ events:
\begin{equation}
N_{bbcc}=N_{4b}\times \frac{\sigma_{bbcc}}{\sigma_{4b}}\times \left(\frac{\epsilon^\mathrm{tag}_c}{\epsilon^\mathrm{tag}_b}\right)^2,
\end{equation}
where $N_{4b}$ is the estimated number of QCD $4b$ events, $\sigma_{bbcc}$ and $\sigma_{4b}$ are parton level cross sections for $bbcc$ and $4b$ processes, $\epsilon^\mathrm{tag}_c$ and $\epsilon^\mathrm{tag}_b$ are the $b$-tagging efficiencies for $c$-quark and $b$-quark jets taken to be 0.2~\cite{Aad:2015ydr} and 0.7, respectively. The expected number of $bbcc$ background events is two {with a luminosity of 3.2~fb$^{-1}$}, i.e. about 5\% of the total background. The distribution of the reconstructed $4b$ invariant mass for background events is shown in Fig.~\ref{fig:distbg13}. This distribution is obtained after rescaling the momenta of the dijet systems such that their invariant masses are equal to 125 GeV. Good agreement is observed between the background estimate from this analysis and the ATLAS results from Ref.~\cite{Aaboud:2016xco}.

\begin{table}[h]
\centering{\begin{tabular}{|c|c|c|c|c|c|}
\hline
 Backgrounds & $\sigma_{\rm parton}({\rm pb})$ & K factor & Efficiency & Expected yield & ATLAS Ref.~\cite{Aaboud:2016xco}\\
\hline
$4b$ & 287.24 & 1.72 (NLO QCD) \cite{Alwall:2014hca}& 4.02$\times 10^{-5}$ & 37  & 43
\\
\cline{1-6}
$t\bar{t}$ & 72 & 1.60 (N$^3$LO QCD) \cite{Muselli:2015kba} & 1.87$\times 10^{-5}$ & 4.0  & 4.3
\\
\hline
\end{tabular}} \ \ \ 
\caption{Cross section, K factor, and acceptance times efficiency for the different sources of background at the 13 TeV LHC. The expected event yields predicted by the simulation used in this analysis are compared with the expected yields from the ATLAS analysis for 3.2~fb$^{-1}$ at $\sqrt{s}=13$~TeV. The cross-section values in the second column include the K factors listed in the third column.}\label{tab:13comp}
\end{table}

\begin{figure}
  \includegraphics[width=0.6\linewidth]{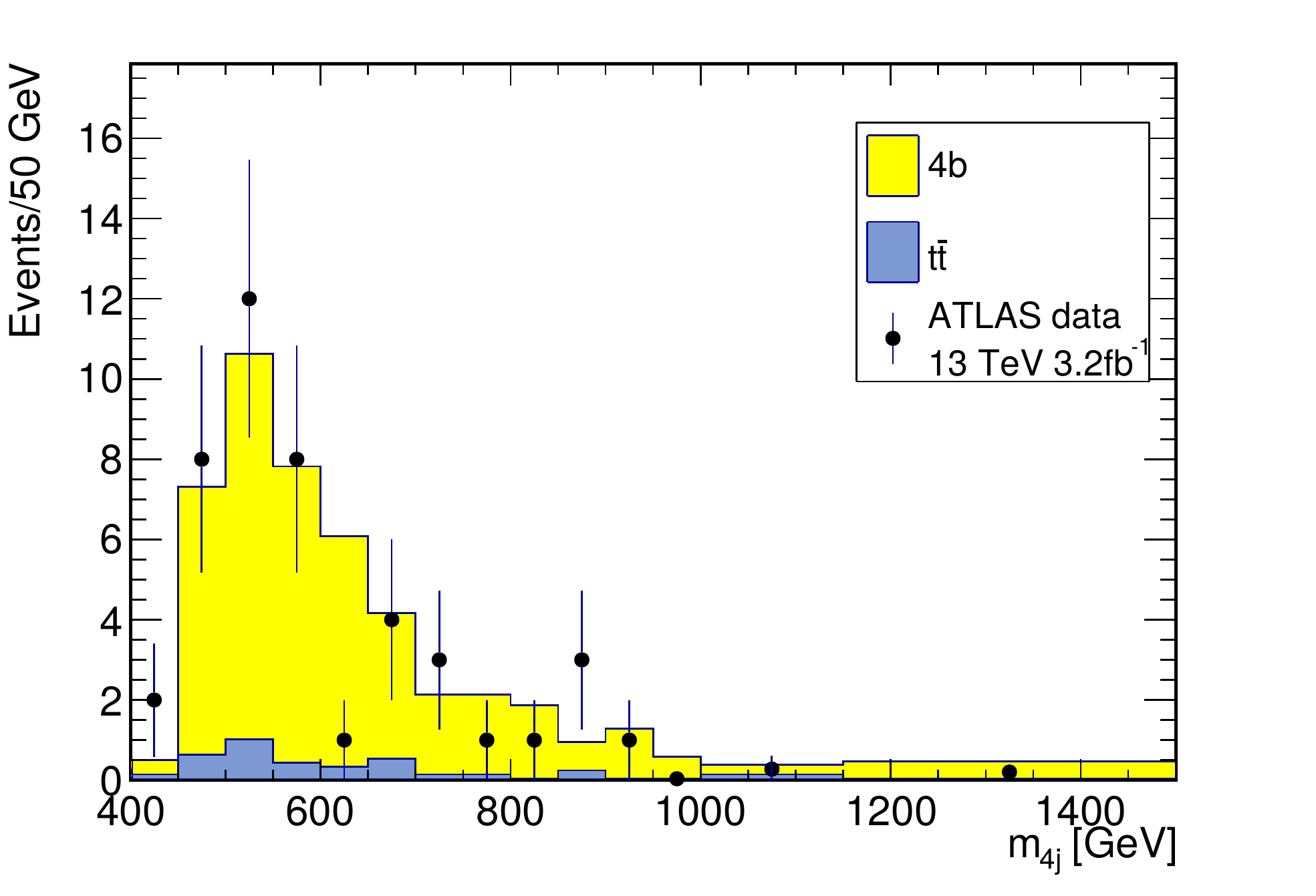}
  \caption{Distribution of the $4b$ invariant mass for events in the signal region. Points correspond to the ATLAS data from Ref.~\cite{Aaboud:2016xco} and histograms to the background sources simulated in this analysis.}
  \label{fig:distbg13}
\end{figure}

\subsection{Predictions for 14 TeV HL-LHC}
After demonstrating that we are able to reproduce the ATLAS results obtained at the 13 TeV LHC,
we evaluate the prospects at the 14 TeV HL-LHC with a modified analysis.
The event selection is modified according to Ref.~\cite{ATLAS:2016ixk}:
\begin{itemize}
\item Events are required to have at least four $b$-tagged jets with $\pt > 30$~GeV and $|\eta| < 2.5$.
\item Dijet systems are formed such that the separation $\Delta R_\mathrm{jj}$ between the two jets satisfies the following requirements:
\[
\left.\begin{array}{r}
\frac{360}{\mfourj}-0.5  < \Delta R_\mathrm{jj}^\mathrm{lead}  < \frac{655}{\mfourj}+0.475 \\
\frac{235}{\mfourj}  < \Delta R_\mathrm{jj}^\mathrm{subl}  < \frac{875}{\mfourj}+0.35 
\end{array}\right\} \mathrm{if}~\mfourj < 1250~\mathrm{GeV},
\]

\[
\left.\begin{array}{r}
0 < \Delta R_\mathrm{jj}^\mathrm{lead} < 1 \\
0 < \Delta R_\mathrm{jj}^\mathrm{subl} < 1
\end{array}\right\} {\rm if}~\mfourj >1250~\mathrm{GeV}.
\]
\item If more than one pair of dijet systems satisfies this constraint, the pair with the smallest variable $D_{h_1h_1}$ is selected with
\begin{equation}
D_{h_1h_1}=\sqrt{(m_{\rm 2j}^{\rm lead})^2+(m_{\rm 2j}^{\rm subl})^2}\left|\sin\left(\tan^{-1}\left(\frac{m_{\rm 2j}^{\rm subl}}{m_{\rm 2j}^{\rm lead}}\right)-\tan^{-1}\left(\frac{115}{120}\right)\right)\right|~.
\end{equation}
\end{itemize}

In order to optimize the separation between signal and background events, the analysis in this paper relies on a BDT trained on half of the simulated signal and background events and validated with the other half. Separate training is performed for each benchmark point studied. The kinematic quantities included in the training of the BDT are
$p_T^{\rm lead},\ p_T^{\rm subl}, \Delta R_\mathrm{jj}^\mathrm{lead}, \Delta R_\mathrm{jj}^\mathrm{subl}, \Delta R_{h_1h_1}, \Delta \phi_{h_1h_1}, \Delta \eta_{h_1h_1}, m^{\rm lead}_{\rm 2j}, m^{\rm subl}_{\rm 2j}, X_{h_1h_1}, m_\mathrm{4j}$, where
the variable $X_{h_1h_1}$ is defined as
\begin{equation}
X_{h_1h_1}\,=\,\sqrt{\left (\frac{\mtwoj^\mathrm{lead}\,-\,120\,{\rm GeV}}{0.1\,\mtwoj^\mathrm{lead}}\right )^2 +\, \left (\frac{\mtwoj^\mathrm{subl}\,-\,115\,{\rm GeV}}{0.1\,\mtwoj^\mathrm{subl}}\right )^2}.
\end{equation}
Among those variables, $\Delta R_\mathrm{jj}^\mathrm{lead}$, $\Delta R_\mathrm{jj}^\mathrm{subl}$, and $m_\mathrm{4j}$ are
consistently ranked high in terms of discrimination power for all benchmark points.
To derive the optimal sensitivity, BDT score distributions for signal and background events are rebinned such that each bin contributes the maximum $S/\sqrt{B}$ ($S$ and $B$ are the numbers of signal and background events in that bin), starting from the bin with the highest BDT score where the signal contributes the most. This rebinning also requires a minimum of ten background events per bin to minimize the impact of statistical fluctuations.
As an illustration, the rebinned BDT score distributions for two benchmark points are shown in Fig.~\ref{fig:bdtBM}.

\begin{figure}
\centering     
\subfigure[~BM4 benchmark]{
\includegraphics[width=0.4\linewidth]{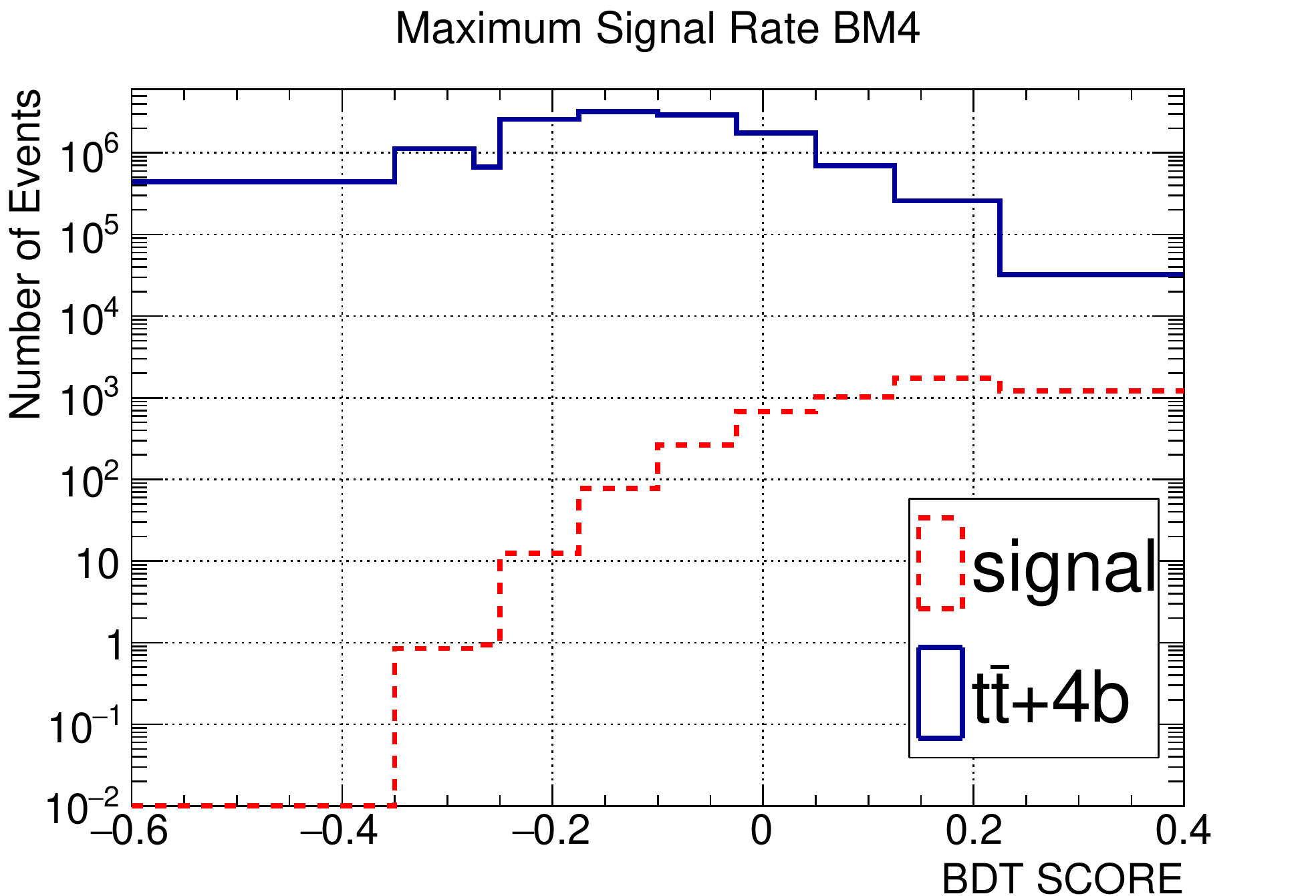}}
\subfigure[~BM7 benchmark]{\includegraphics[width=0.4\linewidth]{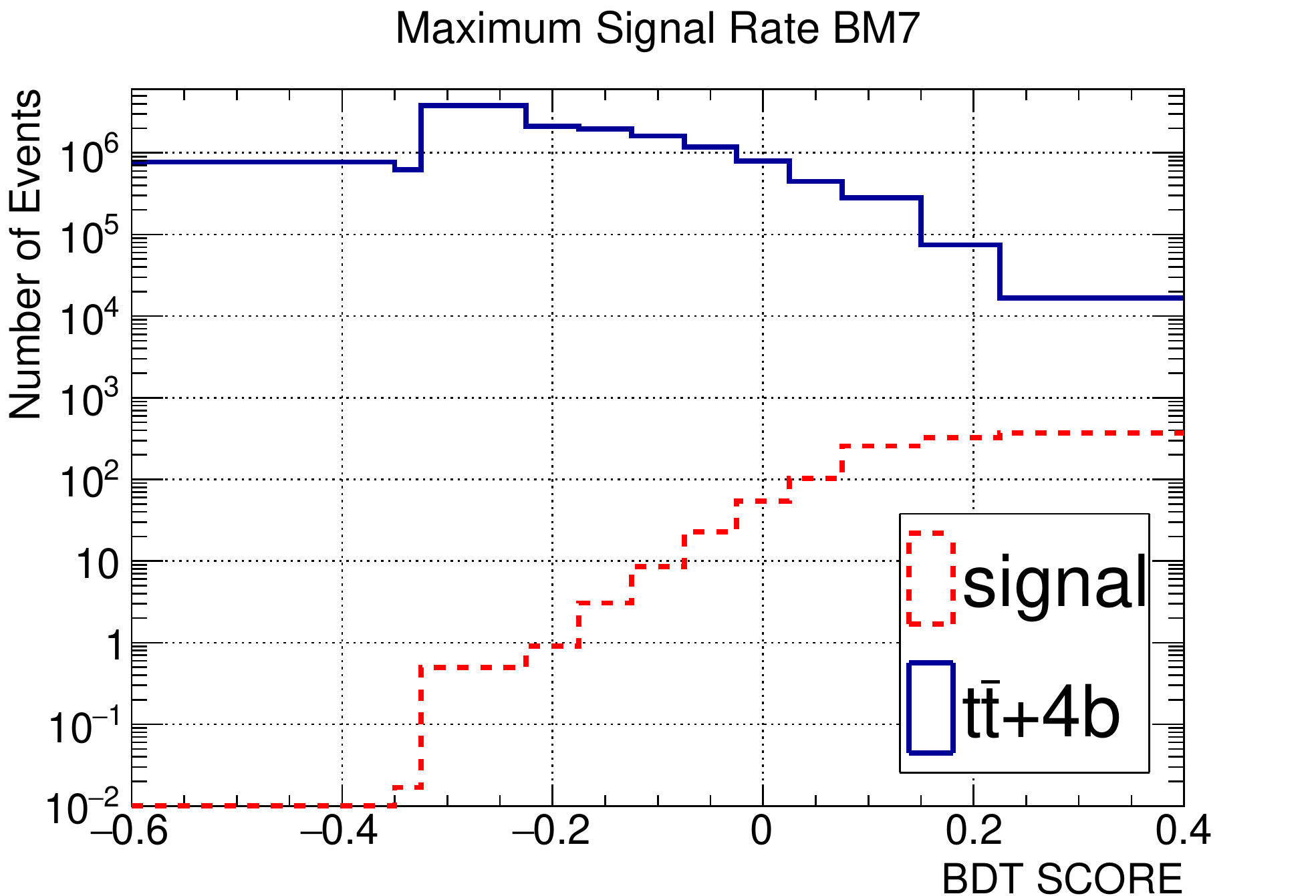}}
\caption{Rebinned BDT score distributions for benchmarks BM4 ($m_2$ = 455 GeV) and BM7 ($m_2$ = 604 GeV) with maximum $S/\sqrt{B}$. The dashed red line represents the signal distribution and the solid blue line represents the background distribution. The distributions are normalized to the expected number of events at the 14 TeV HL-LHC with 3 ab$^{-1}$.}\label{fig:bdtBM}
\end{figure}

The production cross sections and the efficiencies of backgrounds before the BDT selection are summarized in Table~\ref{tab:14BG}. The uncertainties for these backgrounds correspond to the theoretical uncertainties associated with variations in the renormalization and factorization scales and uncertainties from the parton distribution functions.}
In the case of the $pp \to 4b$ background process, a parton-level requirement of $\Delta R_{bb}>0.3$ is imposed in the
generation of events with MadGraph5 to allow the use of the NLO cross-section calculation at $\sqrt{s} = 14$~TeV from Ref.~\cite{Czakon:2015qwa}.
Such a requirement is consistent with the radius parameter $R=0.4$ used in the anti-$k_t$ algorithm as this
sets an effective lower bound of $\Delta R=0.4$ between two jets.

\begin{table}[tbp]
\centering{\begin{tabular}{|c|c|c|c|c|}
\hline
 Backgrounds & $\sigma^{\rm NLO}_{\rm parton}({\rm pb})$ & K factor & Efficiency &Expected yield\\
\hline
$4b$ & ${\rm 130^{+28\%}_{-24\%}}$ & 1.4 (NLO QCD)~\cite{Czakon:2015qwa}& 2.99$\times 10^{-2}$ &$1.17\times 10^7$
\\
\hline
$t\bar{t}$ & ${\rm 110^{+3.8\%}_{-5.8\%}}$ & 2.03 (N$^2$LO +N$^2$LL QCD) \cite{Czakon:2013goa} & 5.58$\times 10^{-3}$ &$1.84\times 10^6$
\\
\hline
\end{tabular}} \ \ \ 
\caption{Cross section, K factor, acceptance times efficiency, and estimated event yields for the different sources of background at the 14 TeV LHC before BDT selection. The cross-section values in the second column include the K factors listed in the third column. Uncertainties in the cross-section values are discussed in the text.}\label{tab:14BG}
\end{table}

To evaluate the sensitivity to di-Higgs scalar resonances, we calculate the ${\rm CL}_b$ value from the rebinned BDT score distributions with the profile likelihood method using the asymptotic formula described in Refs.~\cite{Cowan:2013pha,Cowan:2010js}. The quantity $1-{\rm CL}_b$, which represents the probability that the background-only model yields an observed number of events at least as large as the expectation for the signal plus background model, is then translated into the corresponding $N_{\sigma}$ Gaussian significance.
As a test of the statistical analysis, it was verified that the 95\% upper limit on the cross section as a function of resonance mass derived from our emulation of the 13~TeV ATLAS analysis (discussed in Sec.~\ref{sec:13TeVanalysis}) agrees with the results from Ref.~\cite{Aaboud:2016xco} within 10\% for $h_2$ masses up to 750~GeV and within 20\% up to 850~GeV. The slight deviation at higher mass may be due to the use of the asymptotic formula which is known to produce upper limits that are too aggressive for the low number of expected events at high mass with only 3.2~fb$^{-1}$ of luminosity.

\begin{figure}
\centering     
\subfigure[~BM$_{\rm max}$ benchmark]{
\includegraphics[width=0.4\linewidth]{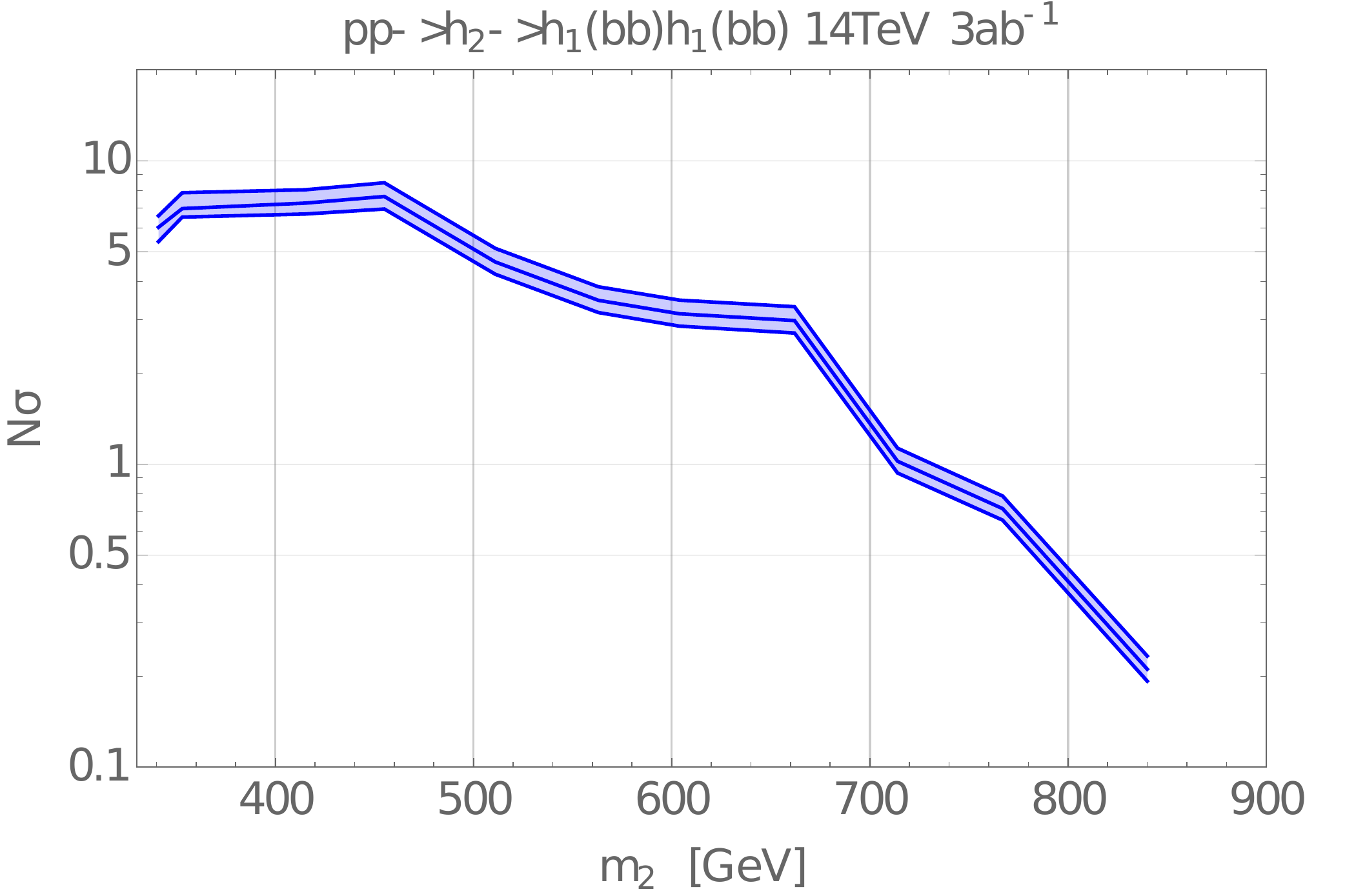}}
\subfigure[~BM$_{\rm min}$ benchmark]{\includegraphics[width=0.4\linewidth]{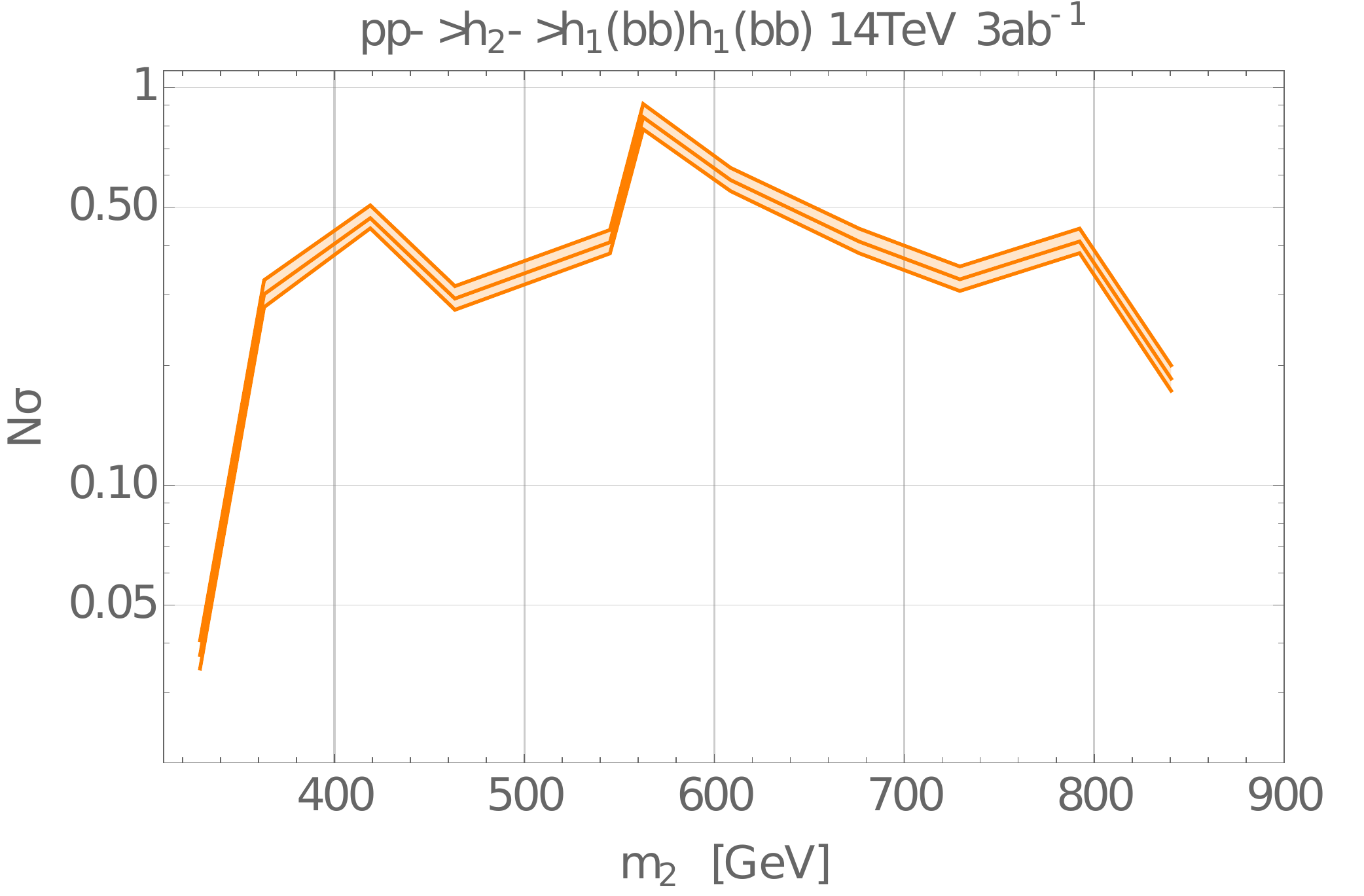}}
\caption{Significance $N_\sigma$ calculated from $1-{\rm CL}_b$ for benchmark models with (a) maximum or (b) minimum cross section in the EWPT scan discussed in Sec.~\ref{sec:EWPT}. The upper and lower bands correspond to the uncertainties in the theoretical cross sections for the $4b$ and $t\bar{t}$ background processes.}\label{fig:14BMsig}
\end{figure}

The significance $N_\sigma$ as a function of resonance mass is shown in Fig.~\ref{fig:14BMsig}, where the upper and lower boundaries of the band correspond to the influence of uncertainties in the production cross sections for the $4b$ and $t\bar{t}$ backgrounds as given in Table~\ref{tab:14BG}. {{The two boundaries are obtained by coherently changing the number of events for the two backgrounds by the 1$\sigma$ uncertainties listed in Table~\ref{tab:14BG}},  computing the CL$_b$ according to the method mentioned above, and then converting  CL$_b$ to $N_\sigma$}. One can observe that, with 3~ab$^{-1}$ of integrated luminosity at the 14 TeV HL-LHC, the benchmark points with maximum signal rate up to $m_2=500$~GeV can be discovered with $N_\sigma>5$. If the future HL-LHC experiments do not observe a signal, then one can exclude the maximum signal rate benchmark points up to $m_2=680$~GeV at 95\% C.L. 

\begin{figure}
  \includegraphics[width=0.6\linewidth]{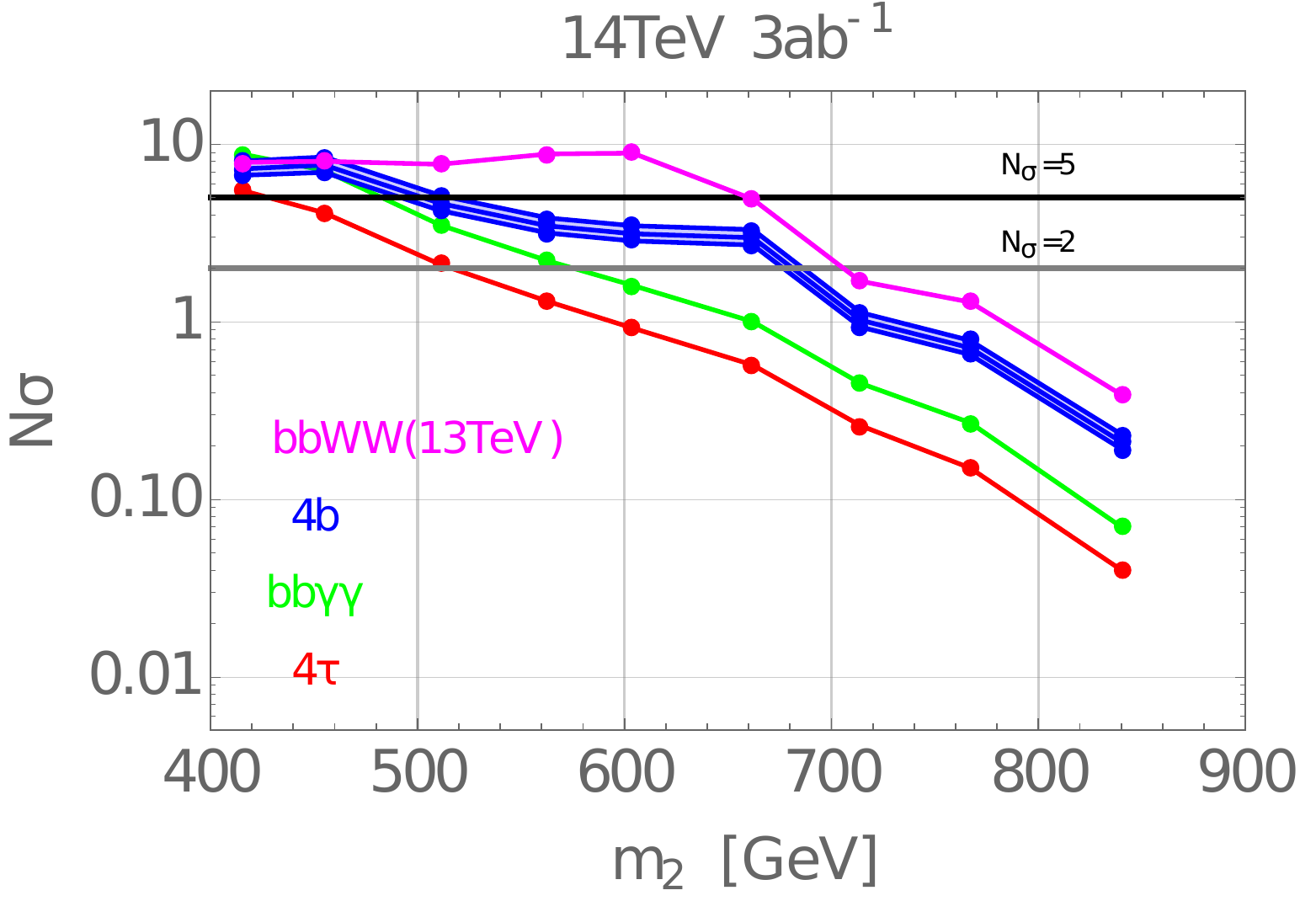}
  \caption{Significance $N_\sigma$ calculated from $1-{\rm CL}_b$ for the 14 TeV HL-LHC with 3~ab$^{-1}$ of integrated luminosity for different channels. {The three blue curves in the plot correspond to the central value and $\pm 1\sigma$ uncertainty bounds obtained by varying the number of background events according to the uncertainties in their cross section (see text).} The values for the $bb\gamma\gamma$ and $4\tau$ channels are obtained from Ref.~\cite{Kotwal:2016tex} whereas those for the $bbWW^\ast$ channel are obtained from Ref.~\cite{Huang:2017jws}.}
  \label{fig:14comp}
\end{figure}

The significance is compared to that obtained with the same method for the $bb\gamma\gamma$ and $4\tau$ channels at the 14 TeV HL-LHC~\cite{Kotwal:2016tex} and for the $bbWW$ channel at the 13 TeV LHC~\cite{Huang:2017jws} in Fig.~\ref{fig:14comp}. 
We only compare the benchmark points from BM3 to BM11 because the first two BM points are different from those in Ref.~\cite{Kotwal:2016tex}. We find that for a heavy Higgs mass $m_2$ less than 500~GeV, the $bb\gamma\gamma$ channel is the most sensitive channel in the search for a resonant di-Higgs signal. Moreover, the $4b$ channel is competitive with the $bb\gamma\gamma$ channel, which could serve as a complementary check if a signal is observed in the $bb\gamma\gamma$ channel. However, for $m_2$ larger than 500~GeV the $4b$ channel provides better sensitivity than the $bb\gamma\gamma$ or $4\tau$ channels but not as good as the $bbWW^*$ channel~\cite{Huang:2017jws}. {We note, however, that the analysis given in Ref.~\cite{Huang:2017jws} employs a novel Heavy Mass Estimator (HME) and assumptions that the systematic uncertainties will be improved compared to those quoted in the recent CMS $bbWW^*$ analysis~\cite{Sirunyan:2017guj} that did not implement the HME. These differences may account for the stronger projected limits given in Ref.~\cite{Huang:2017jws} than one would infer by rescaling the results in Ref.~\cite{Sirunyan:2017guj} by the improved statistics expected for the HL-LHC~\cite{Luca}. We also note that Ref.~\cite{Huang:2017jws} assumes an ATLAS-CMS combination, thereby doubling the number of events. We do not make such an assumption in the present study.}

\section{Conclusion}\label{sec:concl}
{Investigating the thermal history of EWSB is important for determining whether or not the cosmic matter-antimatter asymmetry was generated through EWBG.
Monte Carlo simulations indicate that the EWSB transition is cross over in the minimal SM, given the the observed Higgs mass. In this context of a SM-only universe, EWBG cannot occur.
However, introducing new scalar degrees of freedom can change the behavior of thermal effective potential and make the SFOEWPT possible during the EWSB era. 
Adding a real scalar singlet is one of the simplest ways to extend the SM -- yielding the xSM -- and realize this possibility.
Previous studies have demonstrated the existence of a strong correlation between an enhanced coupling of a heavy singlet-like scalar to a SM-like di-Higgs pair and the occurrence of a SFOEWPT in the xSM parameter space. 
Therefore, there exists strong motivation to search for resonant production of heavy singlet-like scalar that decays to SM-like di-Higgs state as a probe of SFOEWPT in the xSM.

In this paper, we focused on the possibility of discovering at the HL-LHC a resonant gluon fusion production of the heavy singlet-like scalar in the xSM that decays into a pair of SM-like Higgs with a four $b$-quark final state. 
The four $b$-quark final state is a promising  channel, given its large branching ratio,  but it also suffers from a significant QCD background. In analyzing this process, we first validated our simulation against the ATLAS 13 TeV cut-based analysis, then implemented the BDT, a multi-variable analysis method to help to classify signal and background events for the HL-LHC. }
{We selected 11 benchmark points for both maximum and minimum di-Higgs signal rates that yield a SFOEWPT and that satisfy  all the theoretical and phenomenological bounds for a heavy singlet-like scalar mass in successive 50 GeV energy bins ranging from 300 to 850 GeV. We then analyzed the signal significance for the 14 TeV HL-LHC with a luminosity of 3~ab$^{-1}$. We also compared the results with earlier projections for the $bb\gamma\gamma$ and $4\tau$ channels and find that for the mass of the singlet-like scalar larger than 500 GeV, the significance for the $4b$ channel is superior to both of these other channels.  For heavy singlet-like scalar  mass less than 500 GeV, the significance for the $4b$ state with maximum signal rate can be larger than 5. This significance is comparable to that of the $bb\gamma\gamma$ final state, and is somewhat better than that projected for the $4\tau$ final state. While our projection for the reach using the $4b$ channel is somewhat below that for the $bbWW^\ast$ channel as analyzed in Ref.~\cite{Huang:2017jws}, the latter work utilized a new Heavy Mass Estimator and assumptions about future reductions in systematic uncertainties that await validation with new data. Thus, inclusion of the $4b$ channel in a comprehensive search strategy that also includes the $bb\gamma\gamma$, $4\tau$, and $bbWW^\ast$ channels is strongly motivated. }
In terms of exclusion, we find that for the future 14 TeV HL-LHC, one can exclude the mass of a heavy singlet-like scalar up to around 680 GeV for the benchmark points with maximum signal rate. However, a signal in the case of the minimum signal rate benchmark points is far from being excluded. Therefore in this sense, a future 100 TeV $pp$ collider may be required to fully exclude the possibility of generating SFOEWPT in the xSM.
\section*{Acknowledgement}
HL and MJRM were supported in part under U.S. Department of Energy contract DE-SC0011095. HL is supported by the National Science Foundation of China under Grants No. 11875003. SW is supported in part under U.S. Department of Energy contract DE-SC0010004.

\appendix
\section{Appendix}
\label{sec:bigapp}
\subsection{Single Higgs global fit with ATLAS Run~2 Results}\label{sec:gf}
We use the public code \texttt{Lilith}~\cite{Bernon:2015hsa} to implement a global fit with following observables: 
\begin{eqnarray}
\mu_{X,Y} = \frac{\sigma({X}\to H)\:\mathrm{BR}(H\to Y)}{\sigma^\mathrm{SM}({X}\to H)\:\mathrm{BR}^\mathrm{SM}(H\to Y)} \ .
\end{eqnarray}
Here, $X$ represents the production mode (e.g. gluon fusion, vector boson fusion etc.) and $Y$ represents the final states into which the SM-like Higgs decays. We list the data and the corresponding references we use  from ATLAS Run~2 results in Table~\ref{tab:atlas}. The statistical $\chi^2$ is:
\begin{eqnarray}
\chi^2=(\mu-\mu^{obs})^TC^{-1}(\mu-\mu^{obs}),
\end{eqnarray}
where $C^{-1}$ is the inverse of the covariance matrix $cov[\mu^{obs}_i,\mu^{obs}_j]$. In principle we need to know the whole $n\times n$ covariance matrix ($n$ is the number of observables we use in the global fit) to compute the $\chi^2$, but doing so is impossible, as the full matrix is not provided by the experimental collaboration. Instead, we ignore the off-diagonal part in the covariance matrix and approximate the $\chi^2$ as:
\begin{eqnarray}
\chi^2=\sum_{X,Y}\frac{(\mu_{X,Y}-\mu^{obs}_{X,Y})^2}{\sigma^2_{X,Y}},
\end{eqnarray} 
where $\sigma_{X,Y}$ denotes the $1\sigma$ uncertainty for the given observable. The treatment of asymmetric uncertainties is discussed in the \texttt{Lilith} documentation~\cite{Bernon:2015hsa}. 
\begin{table}[htp]
\caption{Measurements of the single Higgs boson cross section by the ATLAS Collaboration relative to the SM prediction for different production mechanisms as used in the global fit.}
\begin{center}
\begin{tabular}{cccccc}
    \hline
    \hline
          & $\gamma\gamma$ &$\tau\tau$ & $WW^\ast$ & $ZZ^\ast$&  $bb$\\
          \hline
     $ggH$& $0.81^{+0.19}_{-0.18}$\cite{Aaboud:2018xdt} &$1.02^{+0.63}_{-0.55}$\cite{Aaboud:2018pen} & $1.10^{+0.21}_{-0.20}$\cite{Aaboud:2018jqu} & $1.11^{+0.25}_{-0.23}$ \cite{Aaboud:2017vzb}     & N.A.\\
     $VBF$& $2.0^{+0.6}_{-0.5}$\cite{Aaboud:2018xdt} &$1.18^{+0.60}_{-0.54}$\cite{Aaboud:2018pen} & $0.62^{+0.36}_{-0.35}$\cite{Aaboud:2018jqu} & $4.0^{+1.7}_{-1.5}$ \cite{Aaboud:2017vzb}      & N.A.\\
     $VH$& $0.7^{+0.9}_{-0.8}$\cite{Aaboud:2018xdt} &N.A. & N.A. & N.A.     & $1.08^{+0.47}_{-0.43}(WH)$ $1.2^{+0.33}_{-0.31}(ZH)$\cite{Aaboud:2018zhk}\\
     $ttH$& $1.39^{+0.48}_{-0.42}$\cite{Aaboud:2018urx} &N.A. &N.A.& N.A.      &$0.79^{+0.61}_{-0.60}$\cite{Aaboud:2018urx}\\
    \hline
    \hline
\end{tabular}
\end{center}
\begin{center}
\label{tab:atlas}
\end{center}
\end{table}
\bibliographystyle{utphys}

\begin{figure}
  \includegraphics[width=0.6\linewidth]{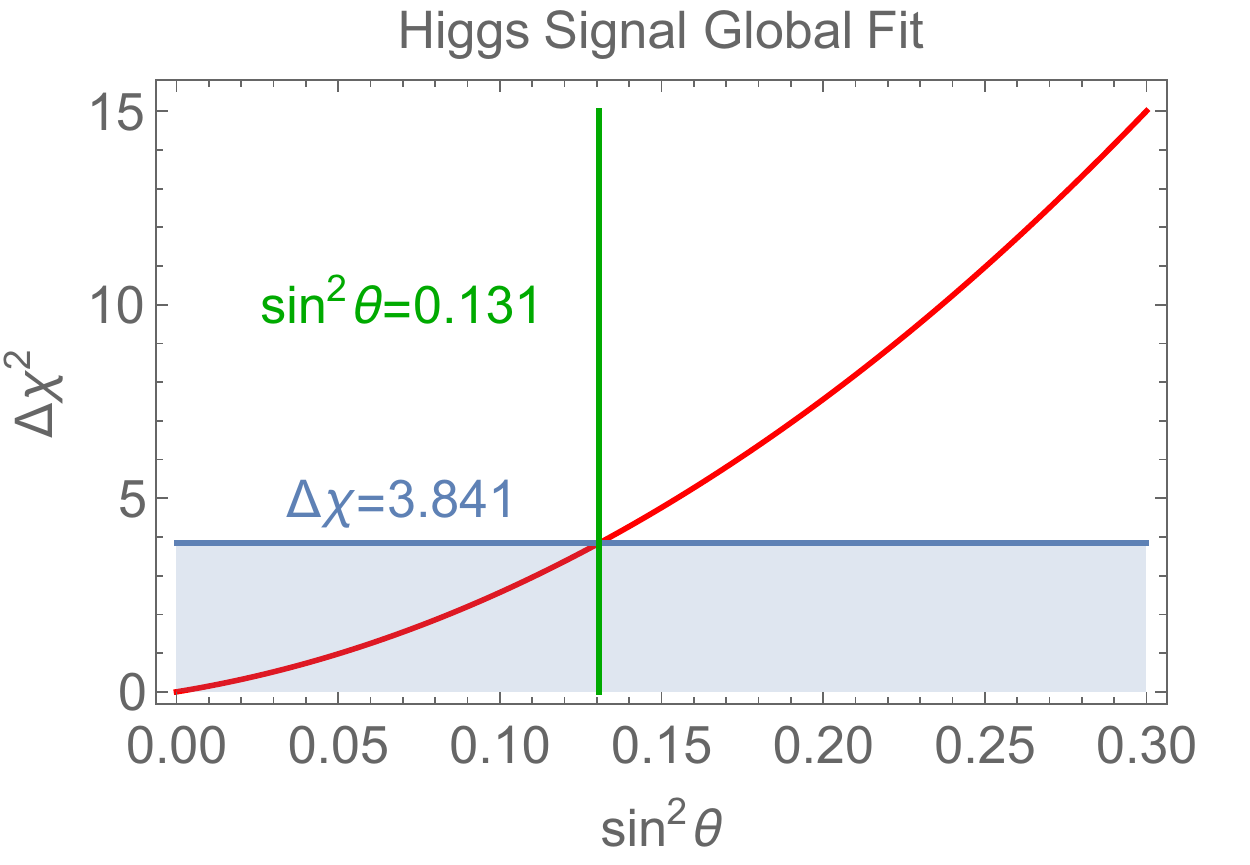}
  \caption{The single Higgs measurements global fit for $\sin^2\theta$ using the ATLAS Run~2 data in Table~\ref{tab:atlas}. The 95\% CL upper limit corresponds to  $\Delta\chi^2<$3.841.}
  \label{fig:gf}
\end{figure}

Figure~\ref{fig:gf} shows the result of our global fit in the $\Delta \chi^2$ vs $\sin^2\theta$ plane, the $\Delta \chi^2$ is defined by:
\begin{eqnarray}
\Delta\chi^2=\chi^2-\chi^2_{min},
\end{eqnarray}
where $\chi^2_{min}$ is the minimum value of $\chi^2$ in the scan.
This translates into a 95\% C.L. upper bound on $\sin^2\theta<$0.131, given by the requirement that $\Delta\chi^2<$3.841.

\section{Distributions of BDT variables}\label{sec:app}
We plot the signal and background distributions of kinematic variables used in the BDT analysis here. The signal is taken to be the benchmark point B7 in Table~\ref{tab:T1}.

\begin{figure}[htp]
\centering     
\subfigure{\includegraphics[width=0.3\linewidth]{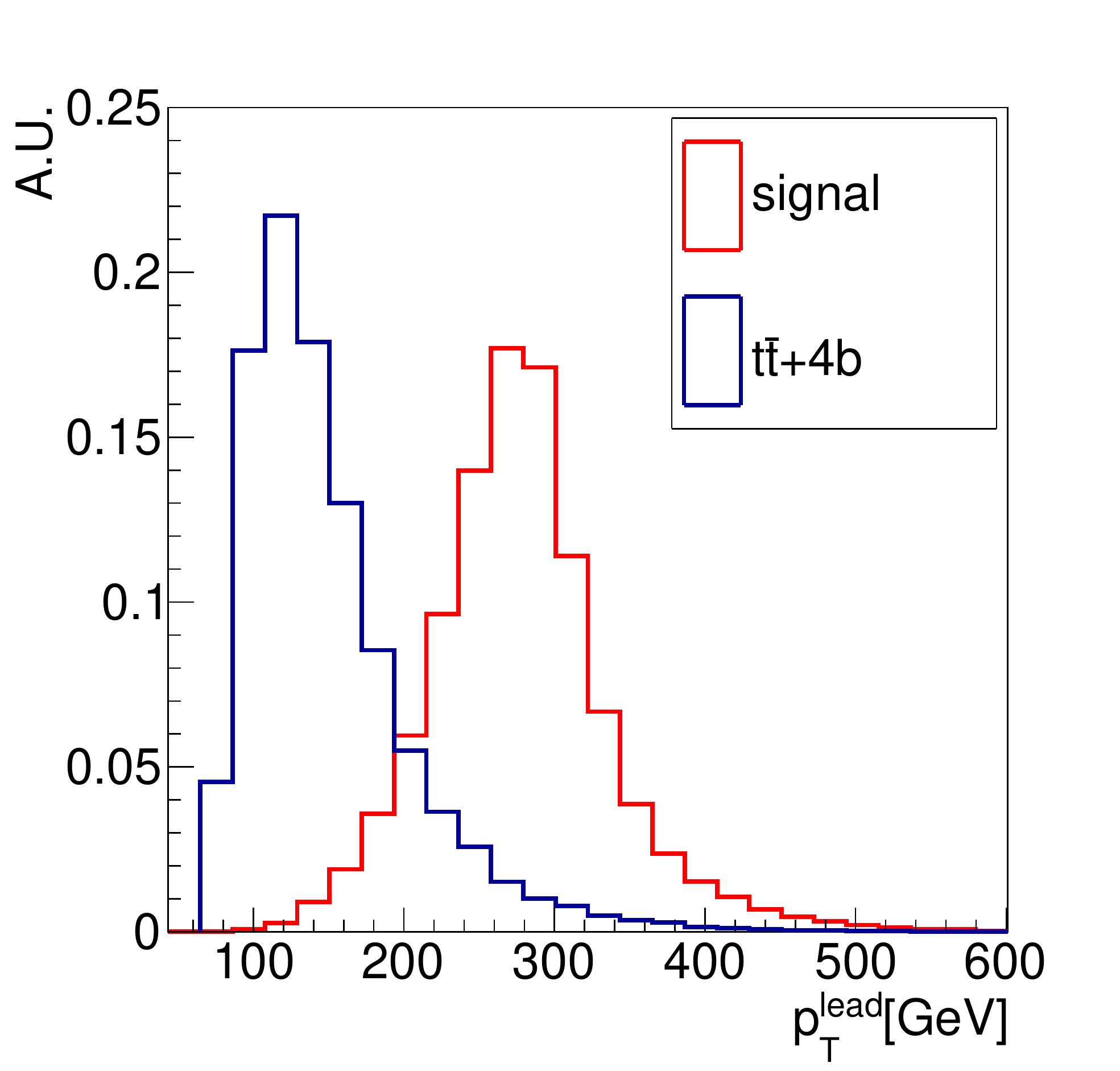}}
\subfigure{\includegraphics[width=0.3\linewidth]{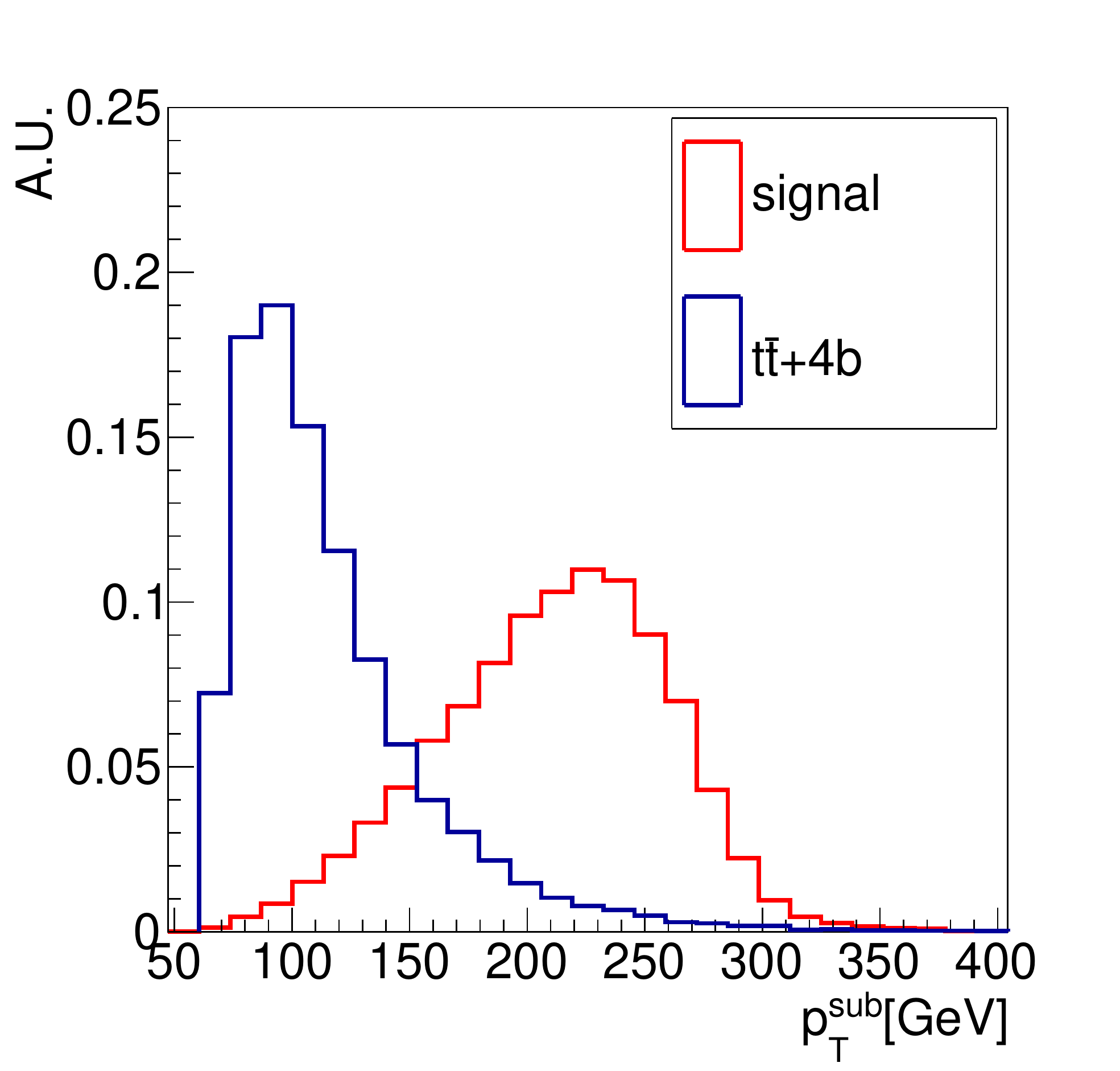}}
\subfigure{\includegraphics[width=0.3\linewidth]{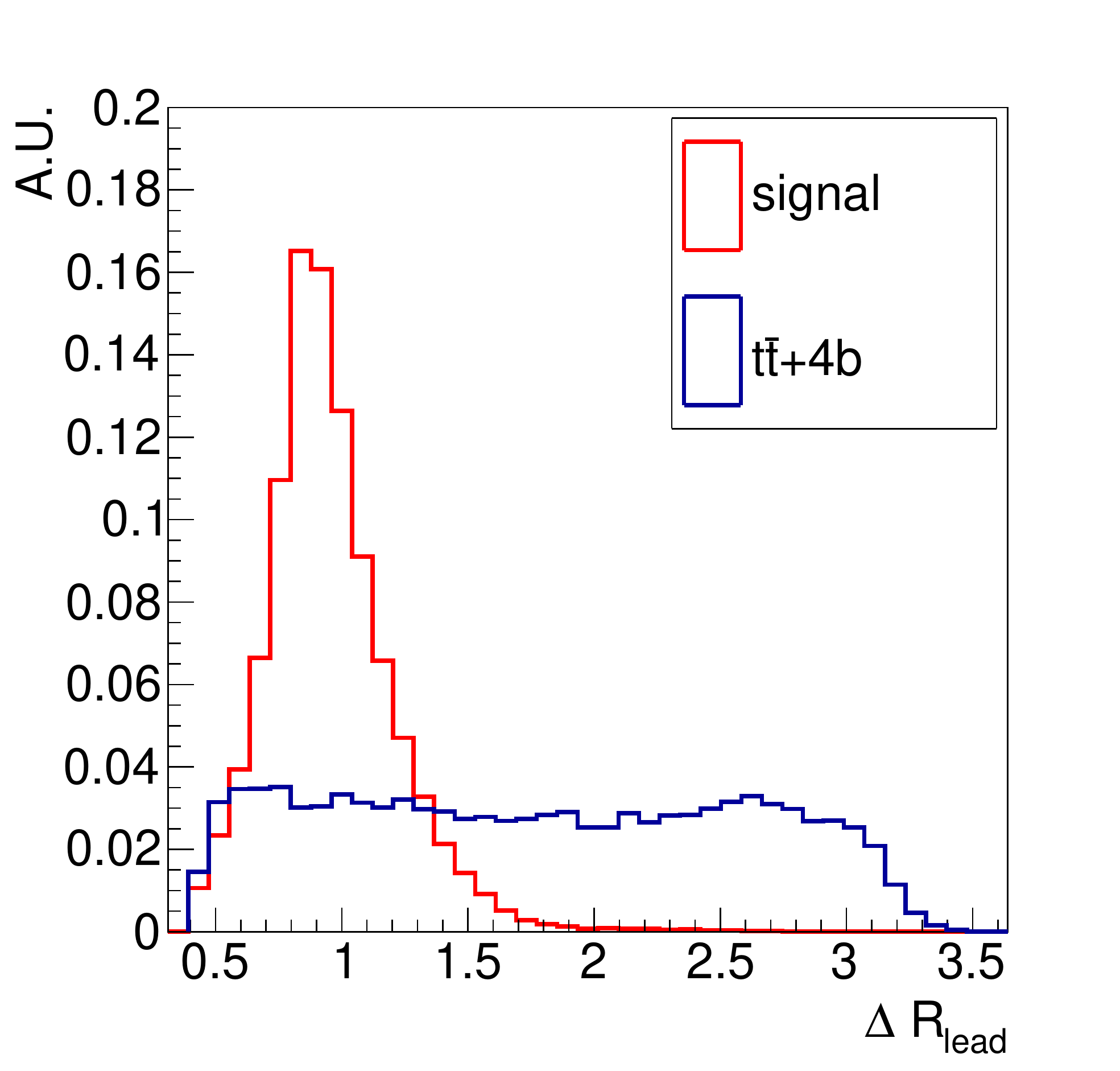}}
\subfigure{\includegraphics[width=0.3\linewidth]{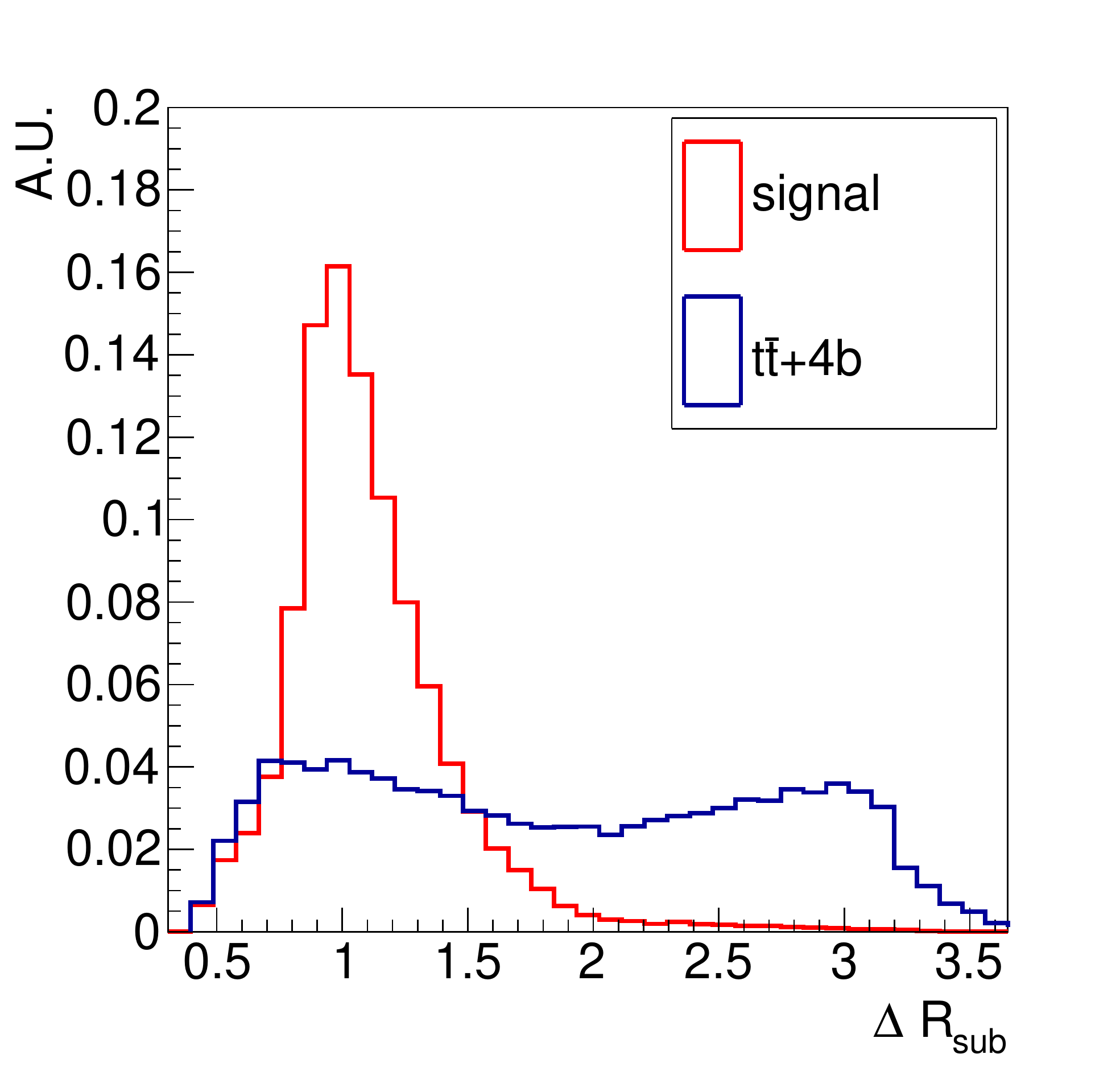}}
\subfigure{\includegraphics[width=0.3\linewidth]{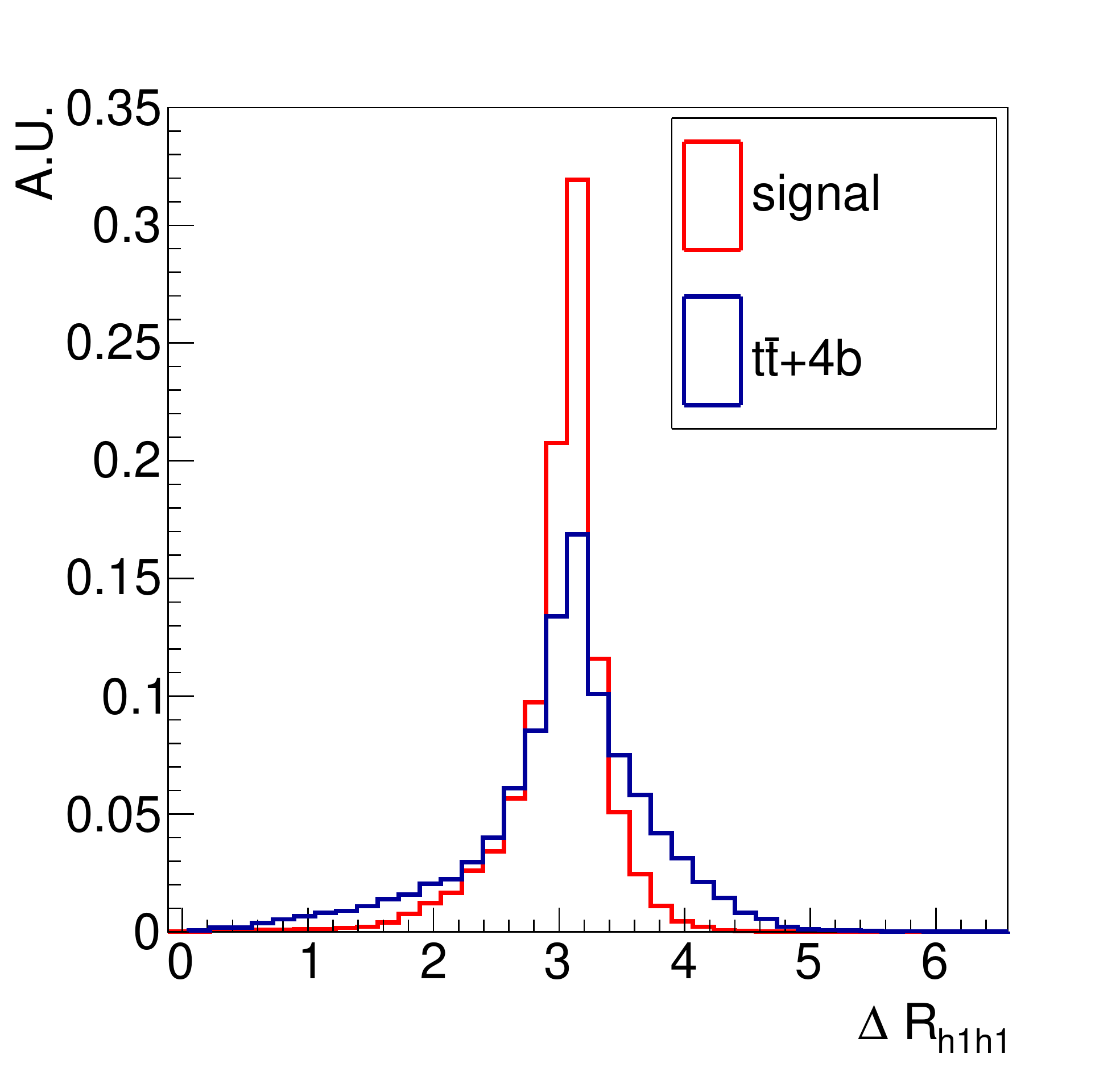}}
\subfigure{\includegraphics[width=0.3\linewidth]{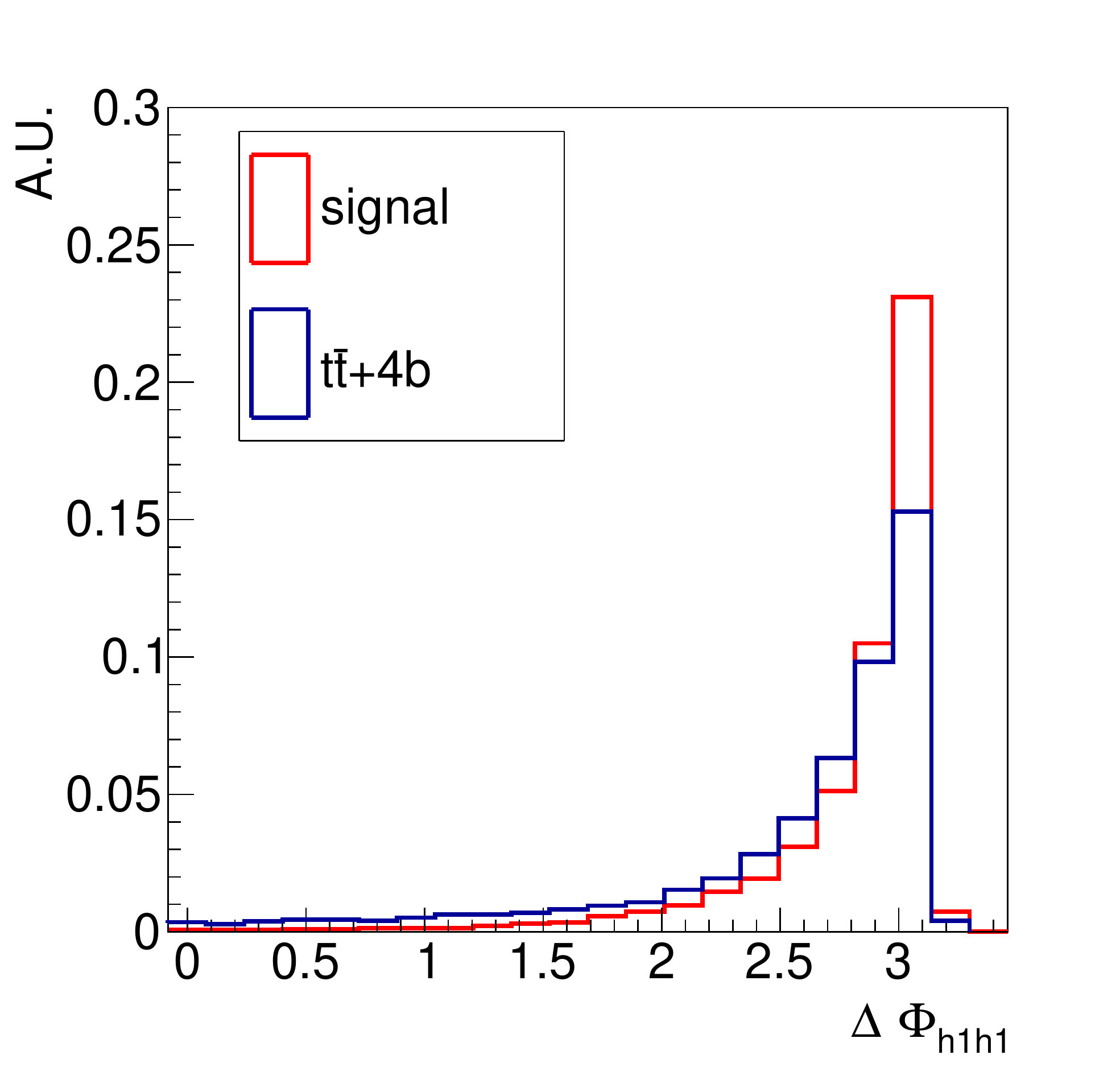}}
\subfigure{\includegraphics[width=0.3\linewidth]{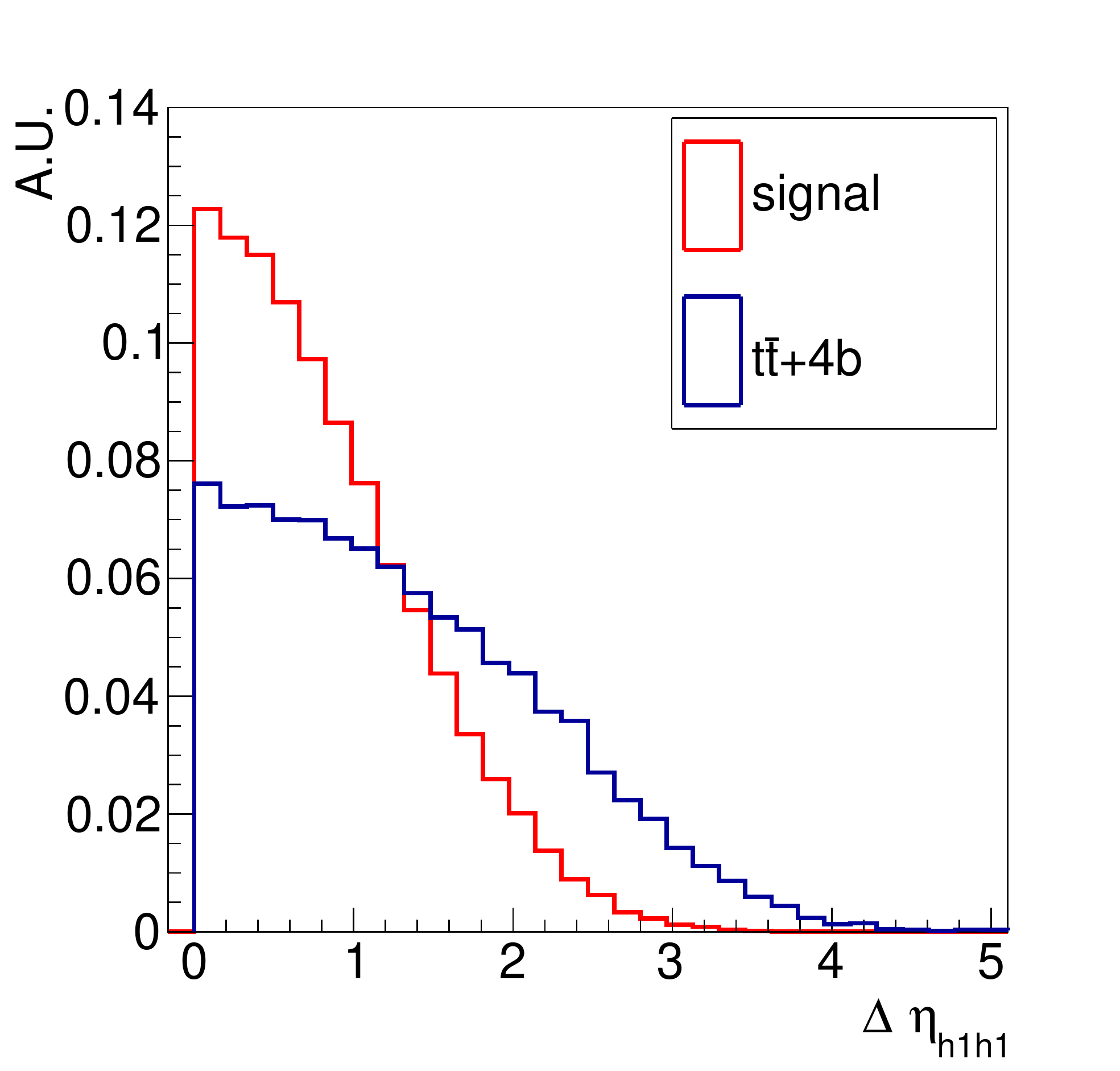}}
\subfigure{\includegraphics[width=0.3\linewidth]{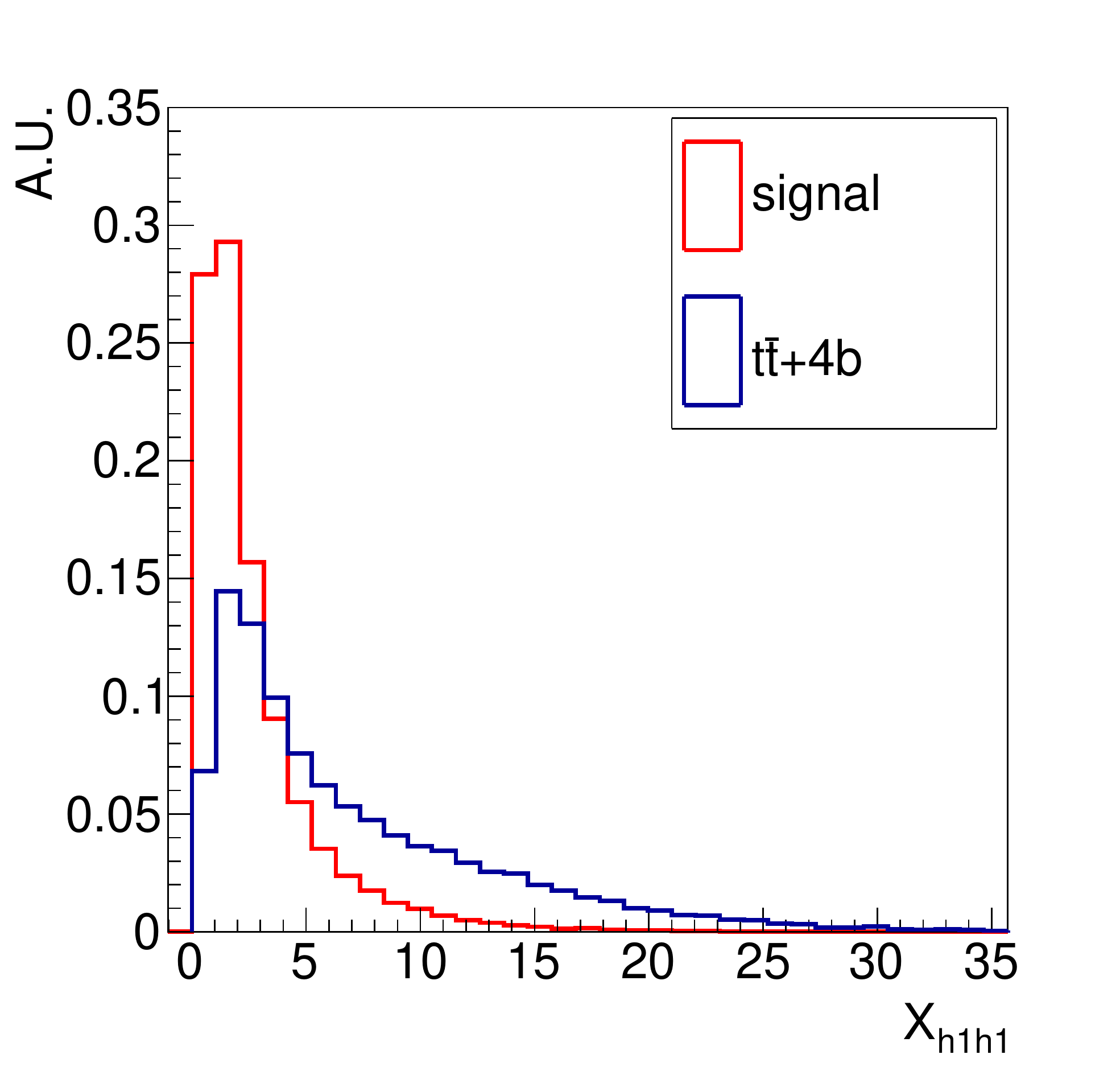}}
\subfigure{\includegraphics[width=0.3\linewidth]{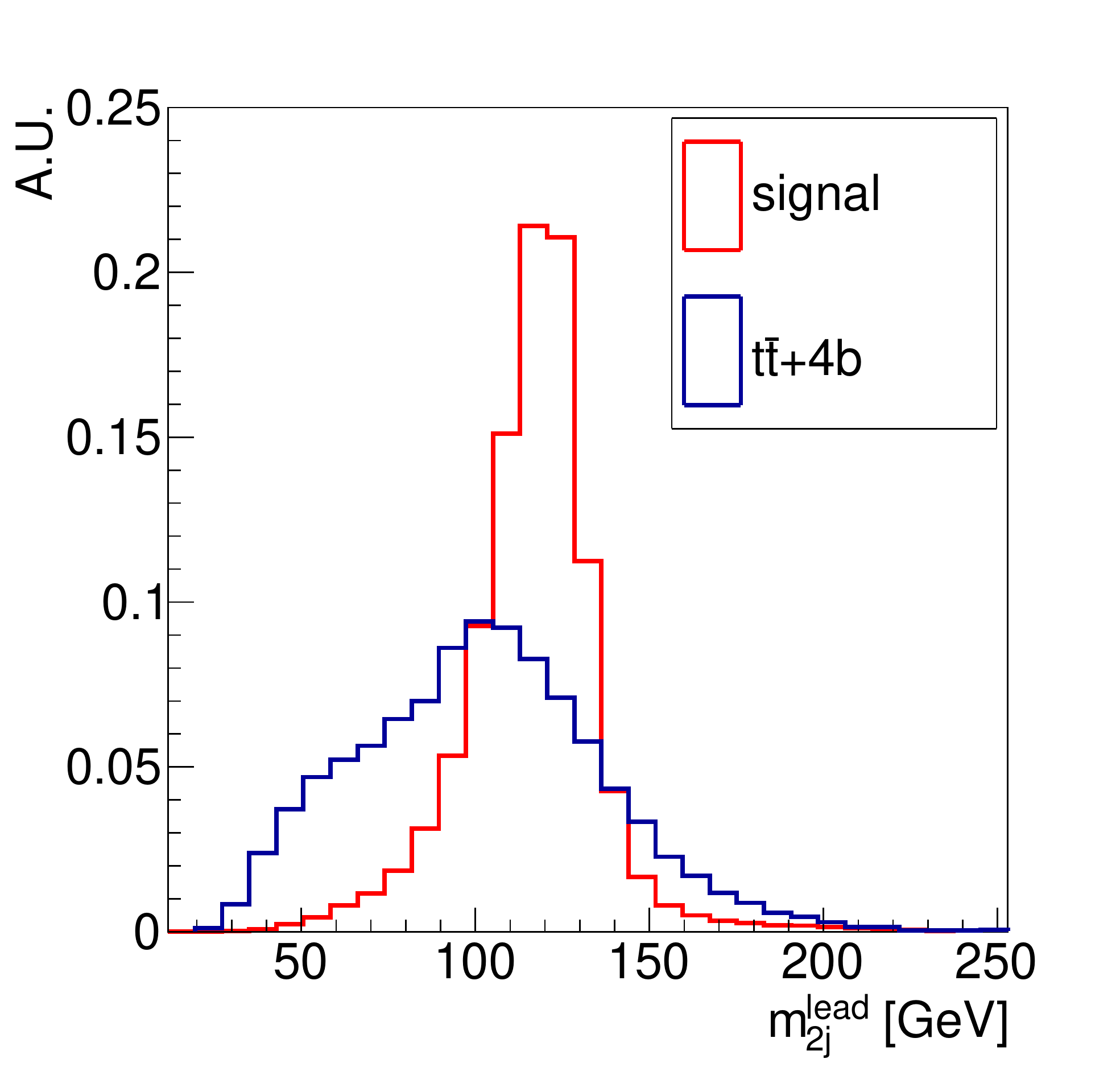}}
\subfigure{\includegraphics[width=0.3\linewidth]{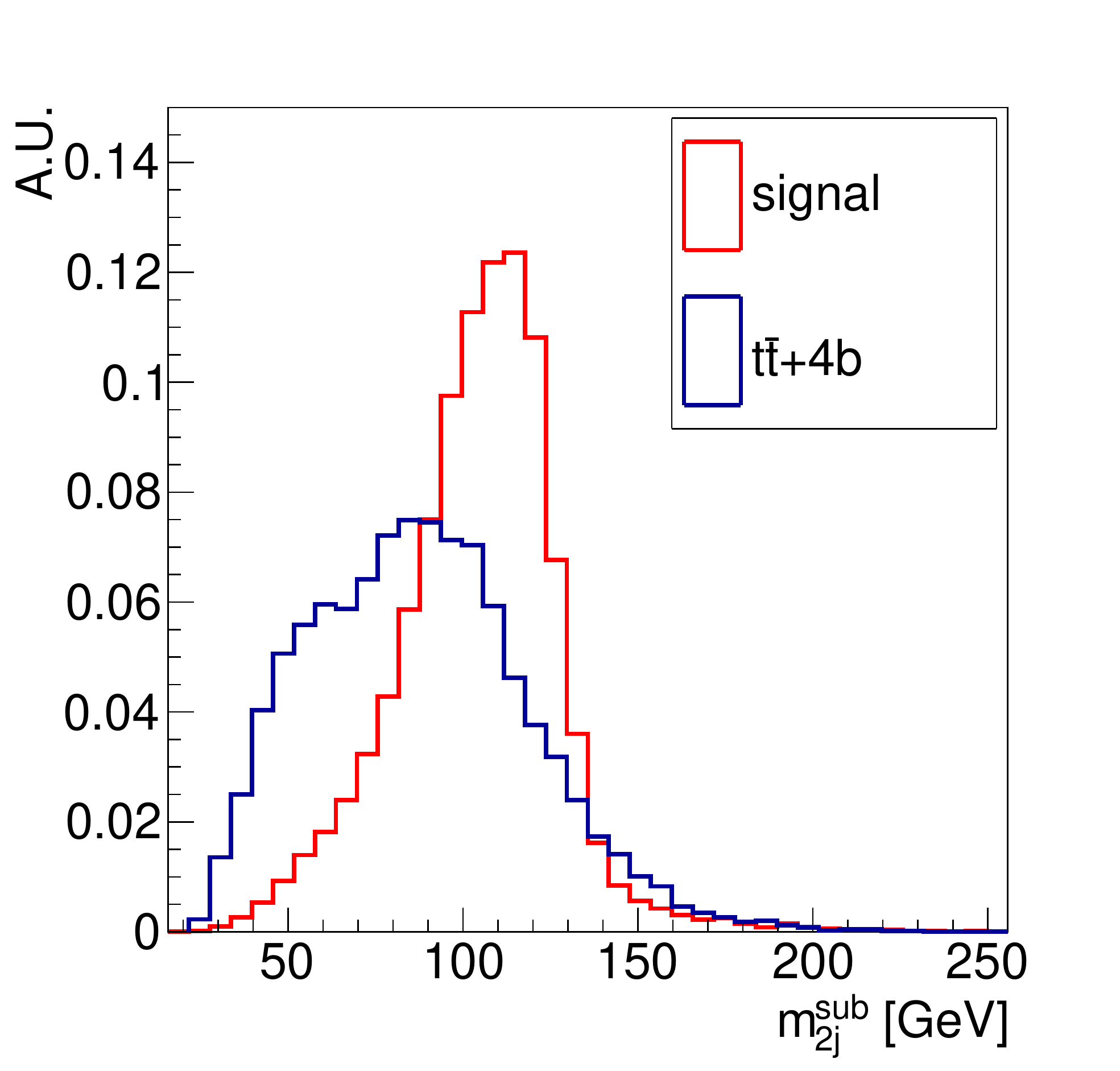}}
\subfigure{\includegraphics[width=0.3\linewidth]{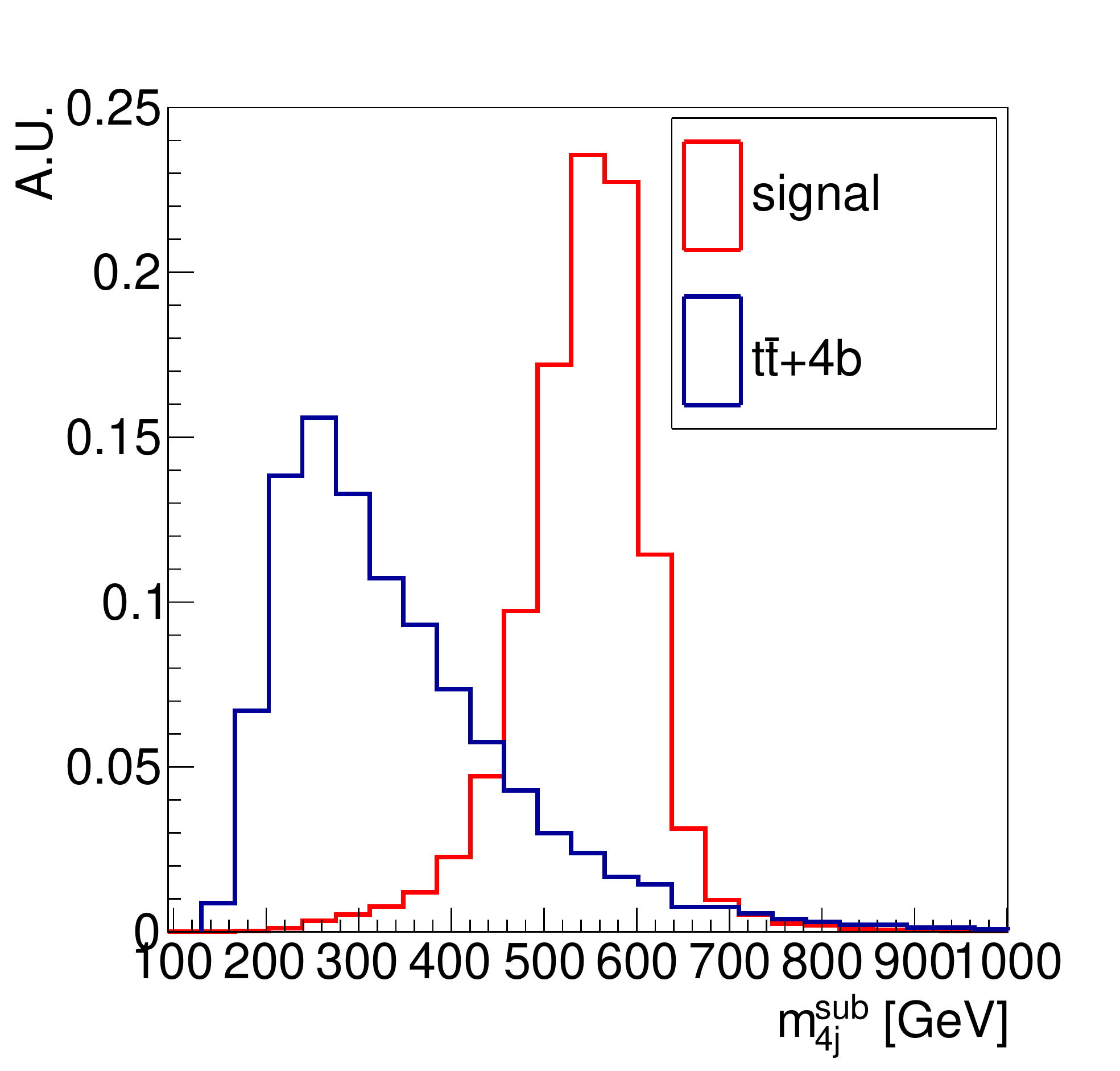}}
\caption{Distributions of the kinematic variables used in training the BDT. The red line represents the signal distribution and the blue line represents the background distribution.}\label{fig:bdtdist}
\end{figure}

\clearpage

\bibliography{bib}

\end{document}